\documentclass[aps,showpacs,amsmath,amssymb,eqsecnum,twocolumn]{revtex4}
\usepackage[T1]{fontenc}
\usepackage{textcomp}
\usepackage[scaled=.92]{helvet}
\usepackage{graphicx}

\begin{document}
\title{Interplay of critical Casimir and dispersion forces}
\author{Daniel Dantchev$^{1,2,3}$\thanks{e-mail:
danield@mf.mpg.de}, Frank Schlesener$^{2,3}$\thanks{e-mail:
schlesen@mf.mpg.de}, and S. Dietrich$^{2,3}$\thanks{e-mail:
dietrich@mf.mpg.de} }
 \affiliation{\mbox{}\\ $^1$Institute of Mechanics -
Bulgarian Academy of Sciences, Academic Georgy Bonchev St. building
4, 1113 Sofia, Bulgaria,\\
$^2$Max-Planck-Institut f\"{u}r Metallforschung,
Heisenbergstrasse 3, D-70569 Stuttgart, Germany,\\
$^3$Institut f\"{u}r Theoretische und Angewandte Physik,
Universit\"{a}t Stuttgart, Pfaffenwaldring 57, D-70569 Stuttgart,
Germany}
\date{\today}

\begin{abstract}

Using general scaling arguments combined with mean-field theory we
investigate the critical ($T \simeq T_c$) and off-critical ($T\ne
T_c$) behavior of the Casimir forces in fluid films  of thickness
$L$ governed by dispersion forces and exposed to long-ranged
substrate potentials which are taken to be equal on both sides of
the film. We study the resulting effective force acting on the
confining substrates as a function of $T$ and of the chemical
potential $\mu$. We find that the total force is attractive both
below and above $T_c$. If, however, the direct substrate-substrate
contribution is subtracted, the force is repulsive everywhere except
near the bulk critical point $(T_c,\mu_c)$, where  critical density
fluctuations arise, or  except at low temperatures and $(L/a) (\beta
\Delta \mu) =O(1)$, with $\Delta \mu=\mu-\mu_c <0$ and $a$ the
characteristic distance between the molecules of the fluid, i.e., in
the capillary condensation regime. While near the critical point the maximal amplitude of the attractive force if of order of $L^{-d}$ in the capillary condensation regime the force is much stronger with maximal amplitude decaying  as $L^{-1}$.  In the latter regime we
observe that the long-ranged tails of the fluid-fluid and the substrate-fluid interactions further increase that amplitude in comparison with systems with short-range interactions only.  Although in the critical region
the system under consideration asymptotically belongs to the Ising
universality class with short-ranged forces, we find deviations from
the standard finite-size scaling for $\xi \ln (\xi/\xi_0^\pm)\gg L$
even for $\xi, L \gg \xi_0^\pm$, where $\xi(t=(T-T_c)/T_c\to\pm
0, \Delta \mu=0)=\xi_0^\pm |t|^{-\nu}$,  is the bulk correlation length. In this
regime the dominant finite-size contributions to the free energy and
to the force stem from the long-ranged algebraically decaying tails
of the interactions; they are not exponentially small in $L$, as it
is the case there in systems governed by purely short-ranged
interactions, but exhibit a power law decay in $L$. Essential
deviations from the standard finite-size scaling behavior are
observed also within the finite-size critical region $L/\xi=O(1)$
for films with thicknesses $L \lesssim L_{\rm crit}$, where $L_{\rm
crit}=\xi_0^\pm (16 |s|)^{\nu/\beta}$, with $\nu$ and $\beta$ as the
standard bulk critical exponents and with $s=O(1)$ as the
dimensionless parameter that characterizes the relative strength of
the long-ranged tail of the substrate-fluid over the fluid-fluid
interaction. We present the modified finite-size scaling pertinent
for such a case and analyze in detail the finite-size behavior in
this region. The standard finite-size scaling behavior is recovered
only for $L\gg L_{\rm crit}$.

\end{abstract}

\maketitle

\section{Introduction}
\label{intro}

\subsection{General properties}

\begin{figure}[htb]
\includegraphics[angle=0,width=\columnwidth]{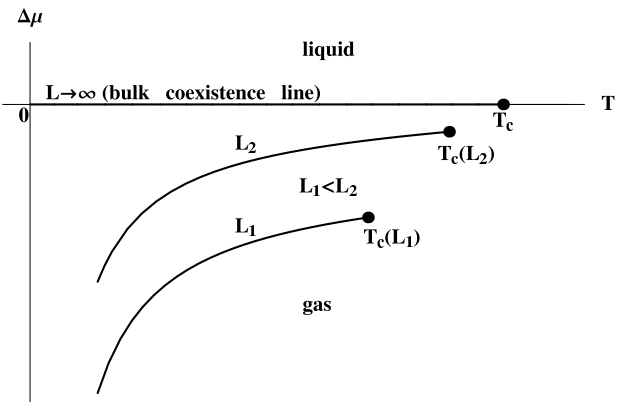}
\caption{The schematic phase diagram of a $d$-dimensional slab with
$(+,+)$ boundary conditions for various thicknesses $L$ and surfaces
which prefer the liquid phase. The bent gas-liquid coexistence
curves correspond to capillary  condensation transitions for $L=L_1$
and $L=L_2$ with $L_1<L_2$ (compare, c.f., Fig. 11(b)).  Away from
the critical region the shift in the phase boundary relative to the
bulk coexistence line $\Delta \mu=0$ is proportional to $L^{-1}$,
while within the critical region it is proportional to
$L^{-\Delta/\nu}$ where $\Delta$ and $\nu$ are the standard bulk
critical exponents. The lines of first-order phase transition  end
at $(d-1)$-dimensional critical points $T_c(L_i)$ with coordinates
$(T_{c,L_i}, \Delta \mu_{c,L_i})$, $i=1,2$, the positions of which
vary with $L$ and depend on the presence   and on the strengths of
the fluid-fluid and the substrate-fluid interactions. For large $L$
these points are located close to the bulk critical point $T_c$ with
coordinates $({T_c,\Delta \mu=0})$: $T_{c,L}-T_c \sim L^{-1/\nu}$
and $\Delta \mu_{c,L} \sim L^{-\Delta/\nu}$.  Since the fluctuations
in systems of reduced size are stronger, one typically has
$T_{c,L_i}<T_c$; the fact that $\Delta \mu_{c,L_i}<0$ expresses the
preferences of the identical walls. \label{pd}}
\end{figure}

Since its first prediction in 1948  by Hendrik Casimir \cite{C48},
the effect named after him has raised significant theoretical and
experimental interest and has been found to occur in numerous
manifestations (see, e.g., Refs.
\cite{MB02,U01,grier02,ZPZ99,Krech,danchev,HSD04,SD06}). Originally,
Casimir considered vacuum fluctuations of the electromagnetic field
between two parallel metal plates which restrict and modify the
fluctuation spectrum  leading   to a dependence of the energy on the
distance $L$ between the plates. This so-called classical (actually
quantum mechanical) Casimir effect, which for decades had been
considered as a theoretical curiosity, in recent years has turned
into a subject of intensive research, not at least triggered by
possible applications in micromechanical devices, and has been
already verified with an impressive experimental accuracy
\cite{lamoreaux,Lam99,BCOR02}. Another manifestation of the Casimir
effect, which has a significant impact towards a different
scientific direction (see, e.g., Refs.
\cite{Krech,FdG78,KD92,GC,ML,UBMCR03,KYP05,danchev,DGS2006,D98,M97,
M99,DKD2003,SHD03,DDG2006,GSGC2006}), is the so-called
statistical-mechanical (thermodynamic) Casimir effect. In a fluid
close to a continuous phase transition at a critical point $T_c$
large fluctuations of the order parameter occur. If, as before, the
fluid is confined by parallel plates at a distance $L$, and is in
contact with a particle reservoir with  a chemical potential $\mu$,
the grand canonical potential $\Omega_{ex}(T,\mu,L)$ of the fluid in
excess to its bulk value $A\, L \, \omega_{\rm bulk}(T,\mu)$ depends
on $L$ so that one can define the effective force $f$ between the
plates per cross sectional area $A$ and per $k_B T$ as
\begin{equation}
\label{def}
f(T,\mu,L)=-\beta\frac{\partial\,\omega_{ex}(T,\mu,
L)}{\partial L}\;,
\end{equation}
where $\beta=1/(k_B T)$, $\omega_{ex}(T,\mu,
L)=\omega(T,\mu,L)-L\,\omega_{\rm
bulk}(T,\mu)=\Omega_{ex}(T,\mu,L)/A$ is the excess grand canonical
potential per cross sectional area $A$, $\Omega(T,\mu,L)=A\,
\omega(T,\mu,L)$ is the total grand canonical potential, and
$\omega_{\rm bulk}(T,\mu)$ is the density of the bulk grand
canonical potential. Besides temperature, chemical potential, and
film thickness the force also depends on which boundary conditions
the surfaces impose on the system. The order near the surfaces can
be either reduced or -- which is the generic case for liquids
confined by solid substrates
 -- increased due to effective surface fields generated by
the confinement. The latter case is known as $(+,+)$ boundary
conditions (for a more precise definition see below). For this case
the schematic phase diagram of a fluid film with thickness $L$ is
shown in Fig. \ref{pd}.

One should keep in mind that the force $f(T,\mu,L)$ is a definition
dependent quantity in the sense that it depends on how one defines
the thickness of the film. A natural choice would be to take the
distance between the planes defined by the positions of the nuclei
of the top layer of each substrate. However, there are certainly
other possible definitions, which will differ by a microscopic
length. This implies that a quantitative comparison between
experimental data and theory is only possible if the data are
accompanied by a precise definition of what $L$ is. Up to now
seemingly there is no awareness of this issue yet.

Close to the critical point, the critical Casimir interactions are
proportional to $k_BT_c$ and therefore the interaction between the
plates can become rather strong in a system with high $T_c$ such as,
e.g., in classical binary liquid mixtures. However, in such systems
the direct dispersion forces between the fluid particles and between
the fluid particles and the substrate particles play also an
important role. The contributions of the dispersion forces to the
total effective force can be distinguished from that of the critical
Casimir forces by their temperature dependence, because the leading
temperature dependence of the former does not exhibit a singularity.
For this reason both in theoretical analyses and in interpretations
of experimental data the contributions due to the dispersion forces
are usually treated separately and, also for the critical region,
are simply added as a regular background contribution to the total
force (see, e.g., Refs. \cite{SHD03,M97,M99}).
\begin{figure*}[htb]
\hspace*{-0.0in}\includegraphics[angle=0,width=2.0\columnwidth]{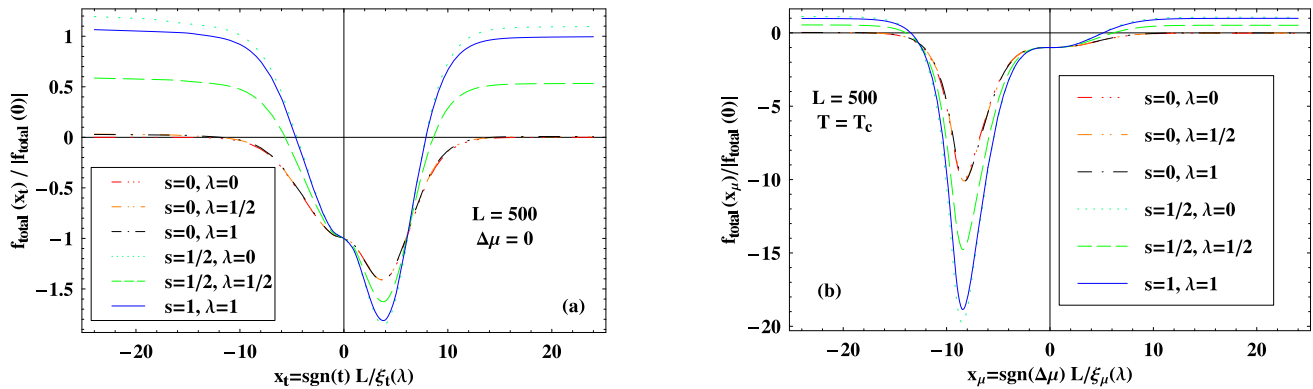}
\caption{$(a)$ Temperature dependence at $\Delta \mu=0$ of the total
Casimir force normalized by its value at the bulk critical point  in
a system with $L=500$ layers for different strengths $\lambda$ and
$s$ of the long-ranged part of the fluid-fluid and fluid-substrate
potentials, respectively; $L$ and $\xi_t$ are measured in units of
the lattice constant of the lattice model. The total force  depends
significantly on the strengths of both potentials. Nonetheless it
turns out that the critical part of the force exhibits universality
and scaling behavior. For small $L$, however, important nonuniversal
contributions due to the long-ranged parts of the aforementioned
interactions emerge which  cannot be neglected. Independent of the
values of $s$ and $\lambda$ the force extremum  always occurs at
positive $t=(T-T_c)/T_c$, i.e.,  {\it above} the bulk critical
temperature. This is due to the stabilizing effect of both the
boundary conditions and of the long-ranged tails of the interactions
on the order of the system so that the strongest fluctuations are
expected to occur slightly above $T_c$ and on the vapor side of the
bulk coexistence curve. Since the Casimir force is generated by
these fluctuations one expects and indeed observes that the
 force is strongest for such thermodynamic parameters. $(b)$
 Same as in $(a)$ for the field dependence $\Delta\mu$ at $T=T_c$.
 The force extremum always occurs at negative values of $\Delta\mu$,
 i.e., on the gas side of the bulk coexistence curve.}
\label{the_total_force_normalized_T}
\end{figure*}

\subsection{Scaling with dispersion forces}
\label{scaling_dispersion}

The present investigation aims at studying in detail the actual
interplay between the dispersion and the critical (fluctuation
induced) forces in simple fluids bounded by strongly adsorbing identical walls. In describing fluid systems near
criticality one often resorts to the Ising model as a representative
of the corresponding universality class, for which only nearest
neighbor interactions are considered. Instead, here we consider
long-ranged pair interactions  between the fluid particles decaying
asymptotically $\sim J^l\, r^{-6}$ for distances $r$ between each
other and long-ranged substrate potentials $\sim J^{l,s}\,z^{-3}$
acting on fluid particles at a distance $z$ from the flat surface of
a semi-infinite substrate. (Here we do not embark into the extended
microscopic description of confined binary fluid mixtures, which are
expected to belong to the same universality class as the
one-component fluids considered here explicitly.) Such systems also
belong to the Ising universality class characterized by short-ranged
forces \cite{Pfeuty}. This implies that the corresponding critical
exponents and leading finite-size dependences  are expected to be
the same as for systems with short-ranged forces.  The long-ranged
part of the interactions leads to a regular contribution to the
force, as considered and expected before, and in addition, as shown
recently \cite{DR2000,D2001,CD2002,DKD2003}, to a serious
modification of the finite-size behavior of its singular
contribution. Indeed, in the case of $d$-dimensional systems, in
which the interaction decays asymptotically with the distance $\sim
r^{-d-\sigma}$ (with $\sigma>2$), the effective total force
$f(T,\mu,L)$ between the plates can be cast (see Eq. (\ref{def}))
into the form \cite{DKD2003,DDG2006,PL83,D86}
\begin{eqnarray}\label{Casimir}
\lefteqn{f(T,\mu,L) \simeq } \\
& & \!\!\!\!\!\!\!\! L^{-d} X[L/\xi_t, L/\xi_\mu,  \left(L/\xi_0\right)^{ -\omega_l}
l, \left(L/\xi_0\right)^{-\omega_s}
s,  \left(L/\xi_0\right)^{-\omega} g_\omega] \nonumber
\\
& & +\beta(\sigma-1)H_A(T,\mu) L^{-\sigma}\xi_0^{\sigma-d}.
\nonumber
\end{eqnarray}
Here $X$ is dimensionless, universal scaling function, $\xi_t$ is
the bulk correlation length $\xi(t\rightarrow \pm 0,\Delta
\mu=0)=\xi_0^\pm |t|^{-\nu}$ at bulk coexistence $\mu=\mu_c$ and for
$t=(T-T_c)/T_c\rightarrow\pm 0$, while $\xi_\mu(t=0,\Delta \mu \to
0)=\xi_{0,\mu}|\beta_c \,\Delta \mu|^{-\nu/\Delta}$ is the bulk
correlation length at the critical temperature $T=T_c$ with $\beta_c=(k_B T_c)^{-1}$. For $T>T_c$
and $\Delta \mu=0$ one has $\xi_0=\xi_0^+$, while for $T<T_c$ and
$\Delta \mu=0$ one has $\xi_0=\xi_0^-$ with the ratio
$\xi_0^+/\xi_0^-$ being universal. $\xi_{0,\mu}$ is the same for
$\Delta \mu \to +0$ and $\Delta \mu \to -0$. The second term in Eq.
(\ref{Casimir}) stems from the free energy contribution $H_A
L^{-(\sigma-1)}$, where $H_A(T,\mu)=A_t(T,\mu)+A_l(T,\mu)
+A_{l,s}(T)$ with so-called Hamaker constants $A_t$, $A_l$, and
$A_{l,s}$    (see below for more details). In addition, $\omega$ is
the standard Wegner's correction-to-scaling exponent for
short-ranged systems, while $\omega_l=\sigma-(2-\eta)$ and
$\omega_{s}=\sigma-(d+2-\eta)/2$ are the correction to scaling
exponents due to the long-ranged tails of the fluid-fluid and
substrate-fluid interactions, respectively. Further $L$-dependent
contributions to the force such as next-to-leading order
contributions to the Hamaker terms or higher order corrections to
scaling are neglected because they are smaller than those captured
in Eq. (\ref{Casimir}). Here we consider a $d$-dimensional system
characterized by an interaction given as the sum of a short-ranged
and a long-ranged component with the latter decaying according to a
power law $r^{-d-\sigma}$, with $2<\sigma<4$. We intend to vary the
ratio $\lambda$ of the strengths of the long-ranged and the
short-ranged contributions. By varying $\lambda$  we can
quantitatively probe the importance of the long-ranged tails. One
might envisage potential experiments in colloidal systems which
allow for a dedicated tailoring of the form of the effective
interactions between colloidal particles. We recall that,
independent of $\lambda$, for  $2<\sigma\leq 4$ the system still
belongs to the corresponding short-ranged universality class, i.e.,
the critical exponents do not depend on $\sigma$. Such long-ranged
interactions are called subleading long-ranged interactions
\cite{DR2000,D2001}. In Eq.~(\ref{Casimir}) $\eta$ is the standard
critical exponent characterizing the decay of the bulk two-point
correlation function at the critical temperature, $g_\omega$ is the (dimensionless) scaling field associated with the Wegner-type corrections, while $l$ and $s$
are dimensionless nonuniversal coupling constants: $l$ is
proportional to the strength $J^l$ of the long-ranged tail in the
fluid-fluid interaction whereas $s$ is proportional to the contrast
between the substrate potential and the fluid-fluid interaction
integrated over a half-space (see below). For the ``genuine''
non-retarded van der Waals interaction, which governs nonpolar
fluids, one has $d=\sigma=3$ and thus $\omega_s>0$, $\omega_l>0$. We
will further suppose that the positivity of $\omega_s$ and
$\omega_l$ is fulfilled for all values of $d$ and $\sigma$
considered in the following. Accordingly, for $L$ \textit{large
enough}, one can expand the scaling function $X$ in Eq.
(\ref{Casimir}):
\begin{eqnarray}
f & \simeq & L^{-d} \Big \{X^{{\rm sr}}\left[L/\xi_t,
L/\xi_\mu\right] \nonumber\\ &&
+ \left(L/\xi_0\right)^{-\omega_s}s
\; \; X_{s}^{{\rm lr}}\left[L/\xi_t,L/\xi_\mu\right]  \nonumber
\\
&&  + \left(L/\xi_0\right)^{-\omega_\sigma} l \; \; X_{l}^{{\rm
lr}}\left[L/\xi_t,L/\xi_\mu \right] \nonumber \\ &&+
\left(L/\xi_0\right)^{-\omega} g_\omega \,X_{\omega}^{{\rm
sr}}\left[L/\xi_t,L/\xi_\mu \right]\Big \} \nonumber \\ && +
\beta(\sigma-1)H_A(T,\mu) L^{-\sigma}\xi_0^{\sigma-d}.\label{manyXf}
\end{eqnarray}

If Eq.~(\ref{manyXf}) is valid the scaling function $X^{{\rm sr}}$,
which originates from the short-ranged interactions and describes
the well studied (short-ranged)  critical behavior, provides the
{\it leading} behavior of the force near the bulk critical point
$(L/\xi_t=0, L/\xi_\mu=0)$. Here the notion ``near the critical
point'' means $L/\xi_t=O(1)$ and $L/\xi_\mu=O(1)$, which defines the
``critical region of the finite system''. There $X_\omega^{\rm sr}, X_l^{\rm lr}$, and
$X_s^{\rm lr}$ represent only {\it corrections} to the leading $L$
dependence. In particular, the value $X^{{\rm
sr}}(0)=(d-1)\,\Delta_{a,b}$ is related to the Casimir amplitude
$\Delta_{a,b}$, which  depends only on the bulk universality class
and the boundary conditions $a,b$ at the two confining surfaces,
i.e., the surface universality classes. There is considerable
knowledge about the Casimir amplitudes, such as their exact values
in $d=2$ \cite{C86,BCN86,P90} and in $d=3$ for the spherical model
\cite{D98}, $\varepsilon$-expansion results for $d=3$
\cite{KD92,DGS2006}, and mean field values \cite{M97}. In the
following we shall consider only the case of $(+,+)$ boundary
conditions corresponding to strong equal surface fields acting at
both surfaces, i.e., the case of strong adsorption of the fluid on
the boundaries of the bounding substrate. While the scaling function
$X^{\rm sr}$ decays exponentially for $L/\xi\gg1$, in this regime
the other scaling functions  $X_{l}^{{\rm lr}}$ and $X_{s}^{{\rm
lr}}$ in Eq.\,(\ref{manyXf}) decay according to a power law. For
this reason these functions, which formally stem from corrections to
scaling due to the subleading long-ranged tail in the fluid-fluid
interaction and in the substrate potential, respectively, lead to
{\it leading finite-size} contributions in the singular behavior of
the force in the regime $L/\xi\gg 1$.  In the case of {\it periodic}
boundary condition  -- in which only the functions $X_l^{\rm lr}$ and
$X_\omega^{\rm sr}$ are present -- the above statement has been verified for
the spherical model (which, for periodic boundary conditions,
represents the limit $n\rightarrow\infty$ of the $O(n)$ models). We note that although  the scaling function $X_\omega^{\rm sr}$ characterizes the short-ranged universality class, the scaling field $g_\omega$ incorporates, in general, also contributions due to the long-ranged tails of the interaction. We refer the interested reader to Ref. \cite{DDG2006} where explicit results about the above mentioned mixing of the  corrections due to the long-ranged forces
and the Wegner-type corrections have been reported. In
the following we are only interested  in the leading $L$-dependence
of the force. Therefore  we shall not discuss in further detail
those contributions which are due to the
corrections to scaling ruled by the Wegner exponent $\omega$
\cite{JJ2002}; they produce corrections \cite{wegner,KR84} both near
the critical point, where they are of the order of $L^{-\omega}$, as
well as away, where they decay exponentially.

\subsection{Relevance-irrelevance criterion}
\label{criterion}

We briefly comment on the conditions which justify an expansion of
the type presented in  Eq. (\ref{manyXf}). The requirements
$2-\eta-\sigma<0$ and $(d+2-\eta)/2-\sigma<0$ are obvious and, as
already mentioned, normally they are satisfied in any realistic
system for which $d=\sigma=3$ and $\eta\ll 1$ (e.g., for the 3d
Ising model $\eta\simeq 0.034$ \cite{JJ2002}). One important
additional condition arises (see Ref. \cite{DR2006}), however, from
the fact that we consider not semi-infinite systems but systems
which are finite in one dimension and thus exposed to power law {\it
long-ranged} substrate-fluid potentials from both sides. Accordingly
these potentials act everywhere in the finite system; at the center
of the film their sum is minimal but not zero. The effect of this
value of the total external potential in the center can be
interpreted as if the system has a nonzero effective  bulk excess
chemical potential $\Delta\mu_{\rm eff}$ despite the fact that the
actual bulk system might be at bulk coexistence curve $\Delta
\mu=0$. Taking into account the contributions from both surfaces in
terms of the notations already introduced one has $\Delta\mu_{\rm
eff}=2 s \left[L/(2\xi_0)\right]^{-\sigma}$. Since the excess
chemical potential scales as $\Delta \mu L^{\Delta/\nu}$ one finds
that in a {\it film} the finite-size contributions due to the
long-ranged substrate-fluid potentials will be negligible {\it in
the critical region} if
\begin{eqnarray}
2 |s| \left[L/(2\xi_0)\right]^{-\sigma}
\left[L/\xi_0\right]^{\Delta/\nu} \ll 1, \end{eqnarray} i.e.,
\begin{eqnarray}
2^{\sigma+1} |s| \left[L/\xi_0\right]^{\Delta/\nu-\sigma} \ll
1.\label{condition}
\end{eqnarray}
The sign of $s$ is chosen such that $s>0\;(<0)$ corresponds to
attractive (repulsive) walls, i.e., walls preferring the liquid
(gas) phase of the fluid. A more detailed discussion of that point
will be given below where we identify $s$ within the framework of a
mean-field model. Due to standard relations  between critical
exponents one has $\Delta/\nu=(d+2-\eta)/2$, so that relation
(\ref{condition}) is consistent with Eq. (\ref{manyXf}). On the
other hand due to $\Delta/\nu-\sigma=d-\sigma-\beta/\nu$ and with
$d=\sigma$ for realistic systems the condition (\ref{condition})
turns into
\begin{equation}\label{condition2}
2^{\sigma+1} |s| \left[L/\xi_0\right]^{-\beta/\nu} \ll 1.
\end{equation}
With $\xi_0$ typically of the order of $3$ {\AA},  $\beta \simeq
0.329$, $\nu \simeq 0.631$ (3d Ising model), and for $\sigma=3$ one
finds
\begin{equation}\label{cL}
L\gg L_{\rm crit} \equiv \xi_0 \left(
2^{\sigma+1}|s|\right)^{\nu/\beta} \simeq 612 \, |s|^{1.918}\;
\mbox{{\AA}}.
\end{equation}
Equation (\ref{cL}) represents a relevance-irrelevance criterion for
van der Waals type substrate potentials in a film of thickness $L$:
if $L\gg L_{\rm crit}$ the leading behavior of the force within the
critical region will be that for a system with short-ranged
interactions, while for thinner films  $L\leq L_{\rm crit}$  the
effects due to the van der Waals substrate potentials are expected
to be relevant. Since $s$ is typically not very small (see below),
$L_{\rm crit}$ turns out to be surprisingly large. As discussed
later for a variety of systems $|s| \in [1,2]$. However,  for some
systems such as $^3$He or $^4$He films  near their bulk liquid-gas
critical point and confined by Au substrate $s$ can be as large as
$4$ \cite{DR2006}. Thus for most fluid systems, for which
measurements of the Casimir force has been performed, $L$ was much
smaller than $L_{\rm crit}$. We stress, however, that until now no
Casimir force measurements for systems with $(+,+)$ boundary
conditions near liquid-gas critical points have been reported.
Therefore we are unable to confront our predictions explicitly with
available experimental data.

Finally, we note that if for a particular choice of the fluid and of the surrounding substrate $s$ happens to be small, also $L_{\rm crit}$, as given by Eq. (\ref{cL}), will be small. Keeping in mind that in Ising-like three-dimensional systems (with $d=\sigma=3$ and $\eta\simeq 0.034$) the correction-to-scaling exponents $\omega \simeq 0.81$ \cite{JJ2002}, $\omega_l \simeq 1.03$  and $\omega_s\simeq0.52$ are numerically not very different from each other, the determination of $L_{\rm crit}$ should in principle in this case take  into account also the values of the scaling fields $g_\omega$ and $l$ in addition to that one of $s$, i.e., Eq. (\ref{cL}) should be modified. However, in such a case an expansion of the  type given in Eq. (\ref{manyXf}) will be valid for any reasonably large $L$ with the contributions due to the Wegner type corrections as well as due to the long-ranged part of the interaction competing with each other and representing corrections to the leading behavior of the force which is given by the scaling function $X^{\rm sr}$.

\subsection{Total Casimir force}

Figure \ref{the_total_force_normalized_T}  presents the typical
behavior of the total (including the contributions from the regular
part of the free energy but neglecting the direct
substrate-substrate interaction) Casimir force as a function of the
temperature or the excess chemical potential, respectively. The
force is normalized by its value at the bulk critical point. The
precise meaning of the parameters $s$ and $\lambda$, as well as the
definition of the model within which the curves have been calculated
will be given below. Larger values of $\lambda$ and $s$ correspond
to stronger  long-ranged tails of the fluid-fluid and
substrate-fluid interactions, respectively. The case ($\lambda=0,
s=0$) corresponds to truly short-ranged interactions. We expect
these curves to resemble potential experimental data for the force
occurring in non-polar fluid films bounded by substrates which
strongly prefer the liquid phase of the fluid. (Experimentally the
contribution of the direct substrate-substrate interaction can be
separated off by a control experiment with the fluid replaced by
vacuum; therefore in the following we do not consider this direct
contribution.) Despite of the spread of the curves for different
values of $s$ and $\lambda$ it will turn out that a description in
terms of scaling functions is possible if the data are suitably
analyzed, provided that for a given value of $s$ the film thickness
$L$ is sufficiently large (see Eq. (\ref{cL})).

It is our aim  to determine the leading $L$-dependence of the
Casimir force as a function of the temperature, the chemical
potential, and of the parameters $s$ and $\lambda$, and to clarify
the relevance of these parameters for its behavior in a fluid film
bounded by strongly adsorbing, identical walls.
To this end and in view of the
large parameter space, starting from a continuum density functional
we describe the system in terms of a lattice gas model which is
expected to capture the essential features of actual fluids as far
as the critical behavior is concerned.  For $L\gg L_{\rm crit}$ the
terms entering into Eq.\,(\ref{manyXf}) are the most important ones
\cite{wegner,JJ2002,KR84}, because for the systems under discussion
here they provide the dominant singular finite-size behavior of the
free energy and of other thermodynamic variables. Whenever $s\ne 0$,
which is the generic or so-called normal case for actual fluids
confined by solid substrates, the functions $X^{\rm sr}$ and
$X_{s}^{{\rm lr}}$ will determine the leading critical finite-size
behavior of the Casimir force. [We note that if the fluid film of
interest is a wetting fluid on a solid substrate or an interfacial
wetting film in binary liquid mixtures, one or two confining
spectator phases are fluids, too. Apart from subtleties such as the
occurrence of capillary waves at confining fluid interfaces, our
present approach is expected to cover the important aspects also of
these more complicated systems. The different preferences of the walls, however, lead to a situation which is not described by $(+,+)$ boundary conditions and requires, therefore, a separate consideration.] Due to the lack of the symmetry $ s
\rightarrow -s$ in the presence  of $(+,+)$ boundary conditions, a
term in the free energy, which is -- to lowest order -- linear in
$s$, is indeed expected. So-called neutral walls, considered
theoretically (see, e.g., Ref. \cite{MEW03} and references cited
therein), correspond to $s=0$. In this special case $X^{\rm sr}$ and
$X_{l}^{{\rm lr}}$ will determine the dominant finite-size behavior
of the fluid system. For such a case $X^{\rm sr}$ and $X_{l}^{{\rm
lr}}$ must be calculated for Dirichlet-Dirichlet boundary
conditions. Recently a similar system of a fluid near a weakly
attractive wall has been considered by Monte Carlo methods
\cite{BGO2004}; but in this study the interaction potential between
the fluid particles has been truncated thereby rendering effectively
a short-ranged potential. Finally, we emphasize again that for film
thickness $L\leq L_{\rm crit}$ an expansion of the type given in Eq.
(\ref{manyXf}) is not possible so that for these systems the van der
Waals interaction will be important throughout the critical region,
including the point $(T=T_c, \Delta \mu=0)$, i.e., in this case even
the determination of the Casimir amplitudes is affected.

\subsection{Outline}

The article is arranged such that in Sec.\,\ref{model} we present
the lattice gas model with van der Waals type long-ranged
interactions. In Sec.\,\ref{results} we present our results for the
free energy and the force by focusing in Subsec.\,\ref{critical} on
our results for  the force near criticality, while in
Subsec.\,\ref{regular} we discuss the behavior of the system off
criticality, including the capillary condensation regime. Finally in
Sec.\,\ref{summary} we summarize our findings and comment on their
experimental relevance. As far as possible we compare our results
with previous findings for fluids governed by or exposed to
dispersion forces. Important technical details are presented in
Appendices \ref{EffectiveHamiltonian}, \ref{offcritical}, and
\ref{dereqstate}.

\section{The model}
\label{model}

Within the density functional approach for inhomogeneous fluids,
which in practice is mean-field-like in character, the grand
canonical functional $\Omega[\rho({\bf r})]$ of a fluid has to be
minimized with respect to the local number density $\rho({\bf r})$
\cite{E79,Di88} (in view of our focus on critical phenomena we
refrain from using more sophisticated versions of density
functionals):
\begin{eqnarray}
\label{omegaDF} \Omega[\rho({\bf r})] &=& \int f_{\rm HS}(\rho({\bf
r}))\\ && + \frac{1}{2}\int\int \rho({\bf r})w({\bf r}-{\bf
r}')\rho({\bf r}')d^3{\bf r}\ d^3{\bf r}' \nonumber \\ && +\int V(z)
\rho({\bf r})d^3{\bf r}-\mu \int \rho({\bf r}) d^3{\bf r}.\nonumber
\end{eqnarray}
The fluid is confined   between two parallel flat plates at a
distance $L$ which exert a substrate potential $V(z)$ with $z$ as
the normal distance from one wall. For an individual wall
$V(z\rightarrow\infty)\sim z^{-\sigma}$ with $\sigma=3$ for a
genuine van der Waals interaction; $\mu$ is the chemical potential
and $f_{\rm HS}(\rho)$ is the bulk free energy density of a
hard-sphere system acting as a reference system; $w(r)$ is the fluid
potential which is suitably regularized at a molecular distance,
i.e., it exhibits   a negative, finite, plateau around $|{\bf r}|=0$
and tends to zero according to a power law for large values of
$r=|{\bf r}|$. In Eq. (\ref{omegaDF}) the integrals run over the
slab volume.

In the spirit of focusing on the essential features of confined
critical phenomena, one can further simplify the continuum
functional in Eq.\,(\ref{omegaDF}) by replacing it by its lattice
version. This resembles the approach taken, e.g., by Fisher and
Nakanishi \cite{FisherNakanishi,Nakanishi-Fisher} in their
mean-field study of the same system but governed by short-ranged
forces. The grand potential functional for this lattice gas system
is
\begin{eqnarray}\label{freeenergyfunctionalstarting}
\Omega[\rho({\bf r})]&=& k_B T \sum_{{\bf r}\in {\cal L}} \Big\{
\rho({\bf r}) \ln\left[\rho({\bf r})\right]\nonumber \\ && +
[1-\rho({\bf r})] \ln\left[1-\rho({\bf r})\right]\Big\}\nonumber \\
&& +\frac{1}{2}\sum_{{\bf r}, {\bf r}'\in {\cal L}}\rho({\bf
r})w({\bf r}-{\bf r}')\rho({\bf r}') \nonumber \\ &&
 +\sum_{{\bf r} \in {\cal L}}[V(z)-\mu] \rho({\bf
r}),
\end{eqnarray}
where ${\cal L}$ is a simple cubic lattice in the region $0 \le z
\le L$ occupied by the fluid. Here and in the following all length
scales are taken in units of the lattice constant $a$ of the order
of a molecular diameter (and thus are dimensionless) so that the
particle density $\rho({\bf r})$ is dimensionless and varies within
the range $[0,1]$. In Eq.\,(\ref{freeenergyfunctionalstarting}) the
terms in curly brackets correspond to the entropic contributions,
while in an obvious way the other terms are directly related to the
interactions present in the system.

The variation of Eq. (\ref{freeenergyfunctionalstarting})  with
respect to $\rho({\bf r})$ leads to the equation of state for the
equilibrium density $\rho^*({\bf r})$
\begin{eqnarray}
\label{eqofstate}
    \lefteqn{2\rho^*({\bf r})-1}\\
    &&=\tanh \Bigg[-\frac{\beta}{2}\sum_{{\bf r}'}
    w({\bf r}-{\bf r}')\rho^*({\bf r}')+
    \frac{\beta}{2}\left(\mu -V(z)\right)
    \Bigg].\nonumber
\end{eqnarray}
The advantage of this equation is that it lends itself to solving it
numerically by iterative procedures. For a given geometry and
surface potential $V(z)$ the solution determines the equilibrium
order-parameter profile $\rho^*({\bf r})$ in the system. Inserting
this profile into Eq.\,(\ref{freeenergyfunctionalstarting}) renders
the grand canonical potential of the system. In order to avoid the
double sum in Eq.\,(\ref{freeenergyfunctionalstarting}), which is
inconvenient for the numerical treatment,  from Eq.
(\ref{eqofstate}) one can easily derive the relation
\begin{eqnarray}\label{asum}
\lefteqn{k_B T \sum_{{\bf r}\in {\cal L}}\rho^*({\bf r})\  {\rm
arctanh[2\rho^*({\bf r})-1]}} \!\!\!\!\!\! \\
&&\!\!\!\!\!\!\!\!\!=\frac{1}{2}\sum_{{\bf r}\in {\cal L}}(\mu -
V(z)) \rho^*({\bf r})-\frac{1}{2}\sum_{{\bf r}, {\bf r}'\in {\cal
L}}w({\bf r}-{\bf r}')\rho^*({\bf r})\rho^*({\bf r}'),\nonumber
\end{eqnarray}
which yields for Eq.\,(\ref{freeenergyfunctionalstarting})
\begin{eqnarray}\label{freeenergyfinal}
\!\!\!\!\!\!\! \!\!\!&& \Omega[\rho^*({\bf r})]=\sum_{{\bf r}\in
{\cal L}}\Bigg[ k_BT\left\{ \rho^*({\bf r}) \ln\left[\rho^*({\bf
r})\right] \right. \nonumber
\\ \!\!\!\!\!\!\! \!\!\! &&  \left.+ [1-\rho^*({\bf r})] \ln\left[1-\rho^*({\bf r})\right]-
\rho^*({\bf r})\ {\rm arctanh[2\rho^*({\bf r})-1]} \right\}
\nonumber \\ \!\!\!\!\!\!\! \!\!\! && -\frac{1}{2}
\left[\mu-V(z)\right] \rho^*({\bf r})\Bigg].\!\!\!\!\!
\!\!\!\!\!\!\!\!\!\!\!\!
\end{eqnarray}
Note that here $\rho^*({\bf r})$ is no longer a free functional
variable, with respect to which one has to minimize
Eq.\,(\ref{freeenergyfinal}), but the solution of
Eq.\,(\ref{eqofstate}).

Denoting $\phi^*({\bf r})=2 \rho^*({\bf r}) -1$ and $\Delta
\mu=\mu-\mu_c$, where $\mu_{c}=\frac{1}{2} \sum_{{\bf r}'} w({\bf
r}-{\bf r}')$, the equation of state (\ref{eqofstate}) can be
rewritten in the standard form
\begin{equation}\label{eqstatestandard}
\phi^*({\bf r})=\tanh \left[ \beta \sum_{{\bf r}'} J({\bf r}, {\bf
r}') \phi^*({\bf r}') +\frac{\beta}{2}\left(\Delta \mu - \Delta
V(z)\right)\right],
\end{equation}
where $J({\bf r}-{\bf r}')=-w({\bf r}-{\bf r}')/4 $. The bulk
properties of the model are well known (see, e.g., Refs.
\cite{D96,B82} and references therein). We recall that the order
parameter $\phi^*$ of the system has a critical value $\phi^*=0$
which corresponds to $\rho_c=1/2$ so that $\phi^*=2(\rho^*-\rho_c)$.
The bulk critical point of the model is given by
$(\beta=\beta_c=[\sum_{\bf r}J({\bf r})]^{-1}, \mu=\mu_c=-2\sum_{\bf
r}J({\bf r}))$ with the sum running over the whole lattice. Within
the mean-field approximation the critical exponents for the order
parameter and the compressibility are $\beta=1/2$ and $\gamma=1$,
respectively. The {\it effective} surface potential $\Delta V(z)$ in
Eq.\,(\ref{eqstatestandard}) is given by (see Appendix
\ref{EffectiveHamiltonian}, in particular Eq. (\ref{hj}) as an
analogue)
\begin{equation}
\label{deltav} \Delta V(z)= \delta v_s
\left[(z+1)^{-\sigma}+(L+1-z)^{-\sigma}\right],
\end{equation}
$1\le z\le L-1$, where contributions of the order of
$z^{-\sigma-1}$, $z^{-\sigma-2}$, etc. have been neglected,
\begin{equation}\label{sdef}
\delta v_s=
-4\pi^{(d-1)/2}\frac{\Gamma\left(\frac{1+\sigma}{2}\right)}
{\sigma\Gamma\left(\frac{d+\sigma}{2}\right)}(\rho_s J^{l,s}-\rho_c
J^l)
\end{equation}
is a  ($T$- and $\mu$-independent) constant,
\begin{equation}
\label{Jfluid} J({\bf r})\equiv J^l_{\rm sr}\left\{\delta(|{\bf
r}|)+\delta(|{\bf r}|-1)\right\}+J^l/(1+|{\bf r}|^{d+\sigma})\;
\theta(|{\bf r}|-1),
\end{equation}
is a proper lattice version of $-w({\bf r})/4$ as the interaction
energy between the fluid particles, and
\begin{equation}
\label{Jsubstrate} J^{l,s}({\bf r})\equiv J^{l,s}_{\rm
sr}\,\delta(|{\bf r}|-1)+J^{l,s}/|{\bf r}|^{d+\sigma}\; \theta(|{\bf
r}|-1)
\end{equation}
is the one between a fluid particle and a substrate particle (here
$\delta(x)$ is the discrete delta function \cite{rem4} and
$\theta(x)$ is the Heaviside step function with the convention
$\theta(0)=0$); $\rho_s$ is the number density of the substrate
particles in units of $a^{-d}$. Note that Eq. (\ref{Jfluid})
incorporates the lattice version of the regularization of $J({\bf
r})$ at ${\bf r}={\bf 0}$ as implied by the original off-lattice
density functional approach. Here we take into account on equal
footing the two contributions to the effective interface potential
\cite{Di88} stemming from the missing fluid particles and from the
substrate particles. From Eq. (\ref{eqstatestandard}) --
(\ref{Jsubstrate}) one can identify the dimensionless coupling
constant
\begin{equation}\label{s_def}
s=-\frac{1}{2}\, \beta\, \delta v_s
\end{equation}
(see Eq. (\ref{manyXf}); $s>0$ corresponds to walls preferring the
liquid phase of the fluid). The case $s=0$, which will be considered
later, corresponds to $\rho_sJ^{l,s}=\rho_c J^{l}$. Note that the
effective potential $\delta v_s$ is formed by the difference of the
relative strength  of the substrate-fluid interaction for a
substrate with density $\rho_s$ and that one of the fluid-fluid
interaction  for a fluid with a density $\rho_c$. In Eq.
(\ref{deltav}) the restriction $z\geq 1$ holds because we consider
the layers closest to the substrate to be completely occupied by the
liquid phase of the fluid (which implies that we consider the strong
adsorption limit), i.e., $\rho(0)=\rho(L)=1$, which is achieved by
taking the limit $J_{\rm sr}^{l,s}\to \infty$; therefore the actual
values of $\Delta V(0)=\Delta V(L)$ will play no role. In order to
preserve the monotonic behavior of $w({\bf r})$ as a function of the
distance $r$ between the particles,  in Eq. (\ref{Jfluid}) we have
to require that $J^l_{\rm sr}\ge J^l/(1+2^{(d+\sigma)/2})$.

In terms of $\phi$ the functional
(\ref{freeenergyfunctionalstarting}) turns into
\begin{eqnarray}\label{gcpomega}
\Omega[\phi({\bf r})]&=& k_B T \sum_{{\bf r}\in {\cal L}} \Bigg\{
\frac{1+\phi({\bf r})}{2} \ln\left[\frac{1+\phi({\bf
r})}{2}\right]\nonumber \\ && + \frac{1-\phi({\bf r})}{2}
\ln\left[\frac{1-\phi({\bf r})}{2}\right]\Bigg\} \nonumber \\ &&
 -\frac{1}{2}\sum_{{\bf r} \in {\cal L}}[\Delta \mu- \Delta V(z)] \phi({\bf
r}) \\
& & -\frac{1}{2}\sum_{{\bf r}, {\bf r}'\in {\cal L}}J({\bf r},{\bf
r}')\phi({\bf r})\phi({\bf r}')+\Omega_{\rm reg},\nonumber
\end{eqnarray}
where
\begin{equation}\label{regomega}
\Omega_{\rm reg}=-\frac{1}{2}\sum_{{\bf r} \in {\cal L}}\left[\Delta
\mu-\Delta V(z)-\sum_{{\bf r}'\in {\cal L}}J({\bf r}, {\bf
r}')\right]
\end{equation}
does not depend on $\phi$ and therefore is a regular background term
which carries a $L$-dependence and thus shows up in the force. An
expression similar to the one in Eq. (\ref{freeenergyfinal}), which
avoids    the double sum and thus is   more convenient for numerical
procedures, can be also obtained. With the identifications
$\phi({\bf r})\leftrightarrow m({\bf
 r})$ and $\frac{1}{2}[\Delta \mu -\Delta V(z)]\leftrightarrow
 h(z)$ one can rewrite the above expression for
 $\Omega[\phi({\bf r})]$ as a
 functional
 $\Delta \Omega[m({\bf r})]\equiv(\Omega-\Omega_{\rm reg})$,
 which describes a magnetic system at temperature $T$ and in the
 presence of an external local and spatially varying magnetic
 field $h(z)$. In the remainder we shall use
 this mutual correspondence between the fluid and magnetic systems
 in order to make contact with existing theoretical results
 for any of them. We shall call the force calculated from using
$\Omega[\phi({\bf r})]$ the {\it total} Casimir force, while that
part of the force calculated with the regular background term
subtracted will be termed the {\it critical} Casimir force. This
procedure corresponds to the analysis of experimental results if one
subtracts from the value of the force measured around $T_c$ the
asymptote obtained by measuring this force well {\it above} $T_c$.

In accordance with Eq.\,(\ref{manyXf}), for the finite-size behavior
of the excess grand canonical potential per unit area $A$ of a
liquid film in the case when both confining surfaces prefer strongly
the liquid phase, one expects
\begin{eqnarray}
&& \label{omega} \omega_{ex}(t,\mu,L|d,\sigma)\simeq
\sigma_{s,1}^{{\rm ns}}+ \sigma_{s,2}^{\rm ns}  \\ && +
k_BT\,L^{-(d-1)} \times \nonumber \\ && X_\Omega[L/\xi_t,(\beta
\Delta \mu) (L/\xi_{0,\mu})^{\Delta/\nu},
(L/\xi_0)^{-\omega_l} l,(L/\xi_0)^{-\omega_s} s] \nonumber \\
&& + \left[A_l(T,\mu)+A_{l,s}(T,\mu)+A_s(T)\right]
L^{-(\sigma-1)},\nonumber
\end{eqnarray}
where $\sigma_{s,1}^{{\rm ns}}$ and $\sigma_{s,2}^{\rm ns}$ are the
non-singular parts of the surface tensions at the surfaces 1 and 2,
respectively (the singular parts are incorporated in $X_\Omega$).
Here
\begin{equation}
\label{at} A_l(T,\mu)=-\frac{4\pi^{(d-1)/2}}{\sigma(\sigma-1)}
\frac{\Gamma(\frac{1+\sigma}{2})}{\Gamma(\frac{d+\sigma}{2})}
J^l\rho_b^2(T,\mu),
\end{equation}
with $\rho_b$ being the bulk fluid density at the given $T$ and
$\mu$, represents that part of the Hamaker constant which is
generated by the long-ranged part of the fluid-fluid interactions,
i.e., the dispersion interaction \cite{rem1,ABUP91}, $A_s<0$ is the
part due to the direct long-ranged interactions between the two
substrates, while $A_{l,s}(T)>0$ is the corresponding term generated
by the long-ranged tails of the substrate potentials acting on the
fluid particles. According to Appendix \ref{EffectiveHamiltonian}
and with $J^{s,s}({\bf r}-{\bf r}')=J^s/|{\bf r}-{\bf
r}'|^{d+\sigma}$ for the interaction between substrate particles one
has
\begin{equation}\label{as}
A_s(T)=-\frac{4\pi^{(d-1)/2}}{\sigma(\sigma-1)}
\frac{\Gamma(\frac{1+\sigma}{2})}{\Gamma(\frac{d+\sigma}{2})}
J^{s} \rho_s^2(T)
\end{equation}
and according to Appendix \ref{Pavel}
\begin{eqnarray}
\label{ah} A _{l,s}(T,\mu) &=& -2 v_s
\rho_b(T,\mu)/(\sigma-1)\\
&=& \frac{8\pi^{(d-1)/2}}{\sigma(\sigma-1)}
\frac{\Gamma(\frac{1+\sigma}{2})}{\Gamma(\frac{d+\sigma}{2})}
J^{l,s}\,\rho_b(T,\mu)\,\rho_s(T)>0,\nonumber
\end{eqnarray}
where (see Appendix \ref{EffectiveHamiltonian})
\begin{equation}
\label{vsd} v_s=-4\pi^{(d-1)/2}
\frac{\Gamma\left(\frac{1+\sigma}{2}\right)}
{\sigma\Gamma\left(\frac{d+\sigma}{2}\right)} \rho_s J^{l,s}.
\end{equation}
Note that $A_l<0$ and $A_s<0$, while $A_{l,s}>0$, and that $A_s$ is
an analytic function of $T$. If both the fluid and the substrates
are governed by van der Waals forces, the constants $A_l$,
$A_{l,s}$, and $A_s$ are not independent. As pointed out in Ref.
\cite{ABUP91} for such systems the parameters $J^l$, $J^{l,s}$, and
$J^{s}$ can be expressed in terms of the corresponding atomic
polarizabilities $\alpha_l$ and $\alpha_s$, i.e.,
$J^l\sim\alpha_l^2$, $J^{s}\sim \alpha_s^2$, and
$J^{l,s}\sim\alpha_l\alpha_s$. (Note that this property implies that
$J^l/\alpha_l^2=J^{s}/\alpha_s^2=J^{l,s}/ \alpha_l \alpha_s$ is
independent of the particular atomic or molecular species.)
Therefore, one can write the sum of $A_l$, $A_{l,s}$, and $A_s$ as a
perfect square, i.e.,
\begin{eqnarray}\label{thesum}
&& A_l+A_{l,s}+A_s =-\frac{4\pi^{(d-1)/2}}{\sigma(\sigma-1)}
\frac{\Gamma(\frac{1+\sigma}{2})}{\Gamma(\frac{d+\sigma}{2})} \times
\nonumber \\ && \left[J^l \rho_b^2(T,\mu)-2J^{l,s} \rho_b(T,\mu)
\rho_s(T)+J^{s}\rho_s^2(T) \right]\nonumber \\ && \sim
 -(\alpha_s \rho_s- \alpha_l \rho_b)^2<0,
\end{eqnarray}
which leads to the conclusion, that  for a Lennard-Jones fluid
between two identical Lennard-Jones walls the effective dispersion
interaction for the film is always attractive. Coating the substrate
surfaces of the system with some additional material does not change
this leading-order $L$-dependence and therefore does not change the
above property. This result is in full agreement with the
Dzyaloshinskii-Lifshitz-Pitaevskii theory \cite{L56,DLP61} which
also predicts that the effective interaction between identical
half-spaces separated by a thin film is always attractive
\cite{ABUP91,AM88}.

The only quantity in Eq. (\ref{omega}), which still has to be
identified for our model, is the value of the coupling constant $l$.
According to Refs. \cite{DDG2006,DR2000,DKD2003}
\begin{equation}
\label{l_def} l=v_\sigma/v_2,
\end{equation}
where $v_\sigma$ and $v_2$ are coefficients in the Fourier transform
$\hat{J}({\bf k})=\sum_{{\bf r}} \exp(i {\bf k}\cdot{\bf r}) J({\bf r})$
of the interaction $J({\bf r})$ (see Eq.(\ref{Jfluid})).   One has
\cite{DKD2003,DDG2006,DR2000,D2001,lattice} $\hat{J}({\bf
k})=\hat{J}({\bf 0})[1-v_2 k^2+v_\sigma k^\sigma-v_4 k^4+O(k^6)]$.
It is easy to check that the short-ranged part $\hat{J}_{\rm
sr}({\bf k})$ of the Fourier transform of the interaction has the
form $\hat{J}_{\rm sr}({\bf k})=J_{\rm sr}^l[1+2d-k^2+O(k^4)]\equiv
\hat{J}_{\rm sr}({\bf 0})[1-v_2^{\rm sr} k^2 +O(k^4)]$, with
$\hat{J}_{\rm sr}({\bf 0})=(1+2d) J^l_{\rm sr}$ and $v_2^{\rm
sr}=1/(1+2d)$ while for the Fourier transform of the long-ranged
part $\hat{J}_{\rm lr}({\bf k})=\hat{J}_{\rm lr}({\bf 0})[1-v_2^{\rm
lr} k^2+v_\sigma^{\rm lr} k^\sigma-v_4^{\rm lr} k^4+O(k^6)]$ an
analytical expression in closed form can be obtained only for the
product $\hat{J}_{\rm lr}({\bf 0}) v_\sigma^{\rm lr}$:
\begin{equation}\label{vs}
  \hat{J}({\bf 0})  v_\sigma =
  \hat{J}_{\rm lr}({\bf 0})  v_\sigma^{\rm lr}
  =J^l n(d,\sigma),
\end{equation}
  with
\begin{equation}
    n(d,\sigma)=-\frac{\pi^{d/2+1}}
    {2^\sigma \sin(\pi \sigma/2)
    \Gamma[(d+\sigma)/2]\Gamma(1+\sigma/2)}>0,
\end{equation}
for $2<\sigma<4$. This leads to
\begin{equation}\label{bdef}
l=\frac{\lambda \; n(d,\sigma)}{1+\lambda \left[\hat{J}^l_{\rm
lr}({\bf 0})v_2^{\rm lr} \right]},
\end{equation}
with
\begin{equation}\label{lambda}
 \lambda=\frac{J^l}{J^l_{sr}},
\end{equation}
while $\hat{J}^l_{\rm lr}({\bf 0})=\hat{J}_{\rm lr}({\bf 0})/J^l$ is
the ground state energy $\hat{J}_{\rm lr}({\bf 0})=J^l \sum_{{\bf
r}}^{\;'} 1/(1+|{\bf r}|^{d+\sigma})$, where ${\bf r}$ with $|{\bf r}|=0,1$ are omitted from the sum (see Eq.
(\ref{Jfluid})), of a system with purely long-ranged interactions
measured in units of $J^l$. Since both $\hat{J}_{\rm lr}(\bf 0)$ and
$v_2^{\rm lr}$ depend also on the properties of the interaction at
short distances it is clear that for the precise determination of
the values of $l$ the use of numerical methods is unavoidable. It
can be shown that
\begin{equation}\label{v2def}
\hat{J}^l_{\rm lr}({\bf 0})v_2^{\rm lr} =\frac{1}{2d} \sum_{{\bf
r}}\frac{|{\bf r}|^2}{1+|{\bf r}|^{d+\sigma}}\; \theta(|{\bf r}|-1).
\end{equation}
For the ``genuine'' van der Waals interaction $d=\sigma=3$ one has
$n(3,3)=\pi^2/12\simeq 0.822$ and if the system is discretized in
terms of a simple cubic lattice one has $\hat{J}^l_{\rm lr}({\bf
0})v_2^{\rm lr}\simeq 1.692$. In this case for $\lambda=0,\, 1/2,\,
1$, and $2$ one has $l=0,\, 0.222, \, 0.306$, and $0.375$,
respectively. Note that for the Lennard-Jones (6,12) potential
\begin{equation}
\label{LJ} w_{\rm LJ}=4 \varepsilon \left[ \left(
\frac{\sigma_0}{r}\right)^{12}-\left(\frac{\sigma_0}{r}\right)^6
\right]
\end{equation}
the position of its minimum, which within our approach can
reasonably be taken as the lattice spacing $a$, is at $a=\sqrt[6]{2}
\;\sigma_0$. The value $-\varepsilon$ of its minimum corresponds to
$J^l_{sr}$. Thus the leading, attractive part $-4\varepsilon
(\sigma_0/r)^6$ of the long-ranged interaction equals $-2\varepsilon
(a/r)^6$, which leads to the conclusion that for pure Lennard-Jones
potentials $\lambda=2$.
\begin{figure*}[htb]
\begin{center}
\begin{tabular}{cc}
\hspace*{-.45cm}\resizebox{0.98\columnwidth}{!}
{\includegraphics{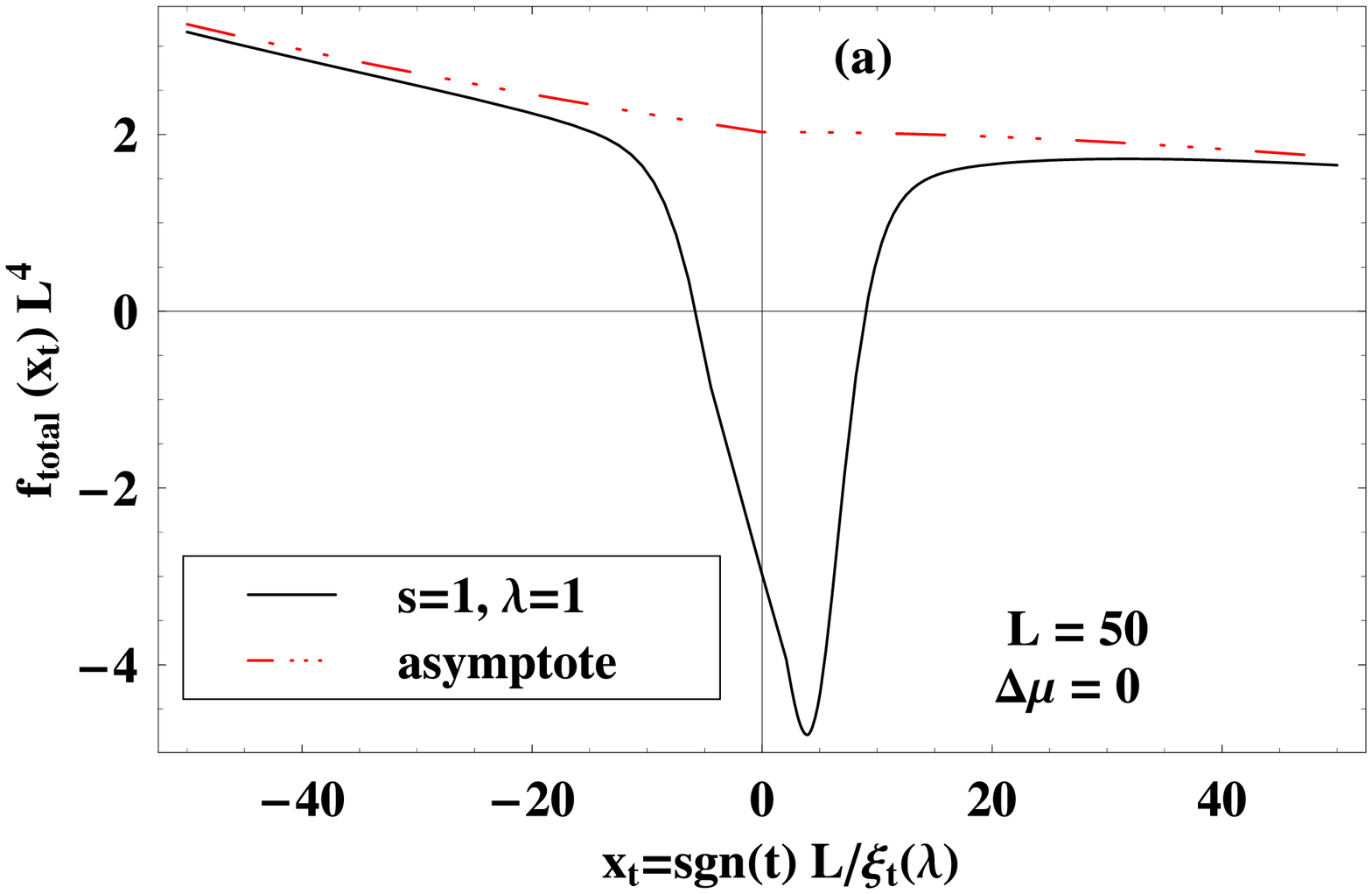}} &
\hspace*{.775cm}\resizebox{0.98\columnwidth}{!}
{\includegraphics{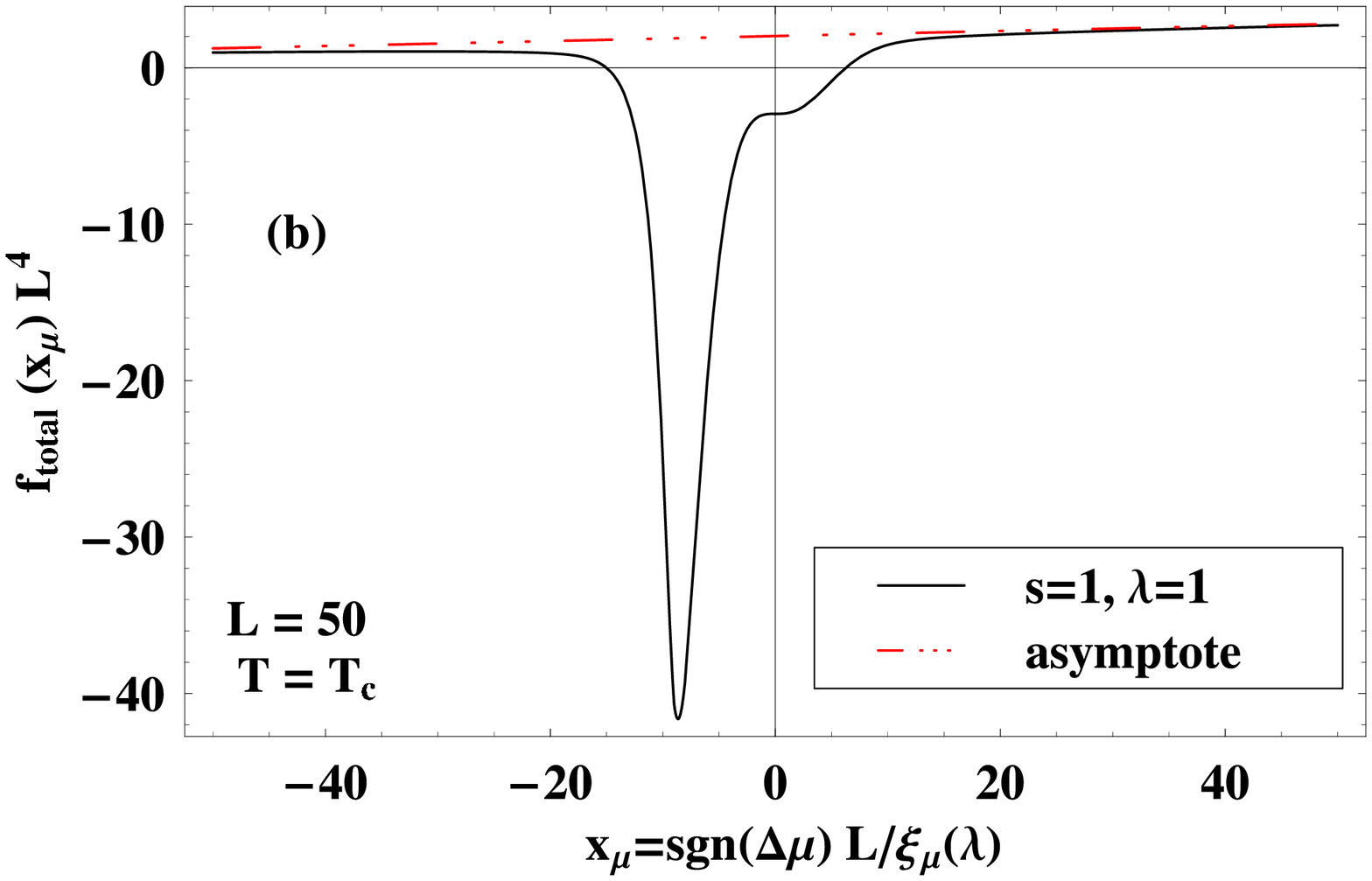}}
\end{tabular}
\end{center}
\caption{$(a)$ The typical behavior of the total Casimir force (per
area of the film cross section, in units of $k_B T$, and without the
direct substrate-substrate interaction) at $\Delta\mu=0$ as a
function of the temperature scaling variable in a system with
substrate-fluid and fluid-fluid long-ranged interactions.  One
observes that  outside the critical region the force is repulsive
and becomes attractive only near the critical point of the system
due to the critical fluctuation of the fluid density. (If the direct
substrate-substrate interaction is added to $f_{\rm total}$, the
resulting force is  attractive  in the whole thermodynamic space.)
The data correspond to $L=50$ and $s=\lambda=1$ (where $s=1$ for
$T=T_c$, see Eq. (\ref{s_def})). The asymptote (see Eqs. (\ref{at})
and (\ref{ah})) is given by $\beta
(\sigma-1)\left(A_{l,s}+A_l\right)=4s K \rho_b+ \lambda K
 \{4\pi^{d-1}\Gamma[(1+\sigma)/2]\}/\{\sigma \Gamma[(d+\sigma)/2]\}
\rho_b \left(1-\rho_b\right)$ where in the present mean-field case
$d=\sigma=4$, $K=\beta J^l_{\rm sr}$, and for $K_c$ see Eq.
(\ref{Kc}). Note that above the critical point $\rho_b=\rho_c=1/2$,
while for $T<T_c$ $\rho_b$ increases upon a decrease of the
temperature. This leads to a clearly visible asymmetry in the
asymptotic behavior of the force. $(b)$ The same as in $(a)$ but as
function of the field scaling variable and at $T_c$.    One again
observes that everywhere outside the critical region the total force
(without the direct substrate-substrate interaction) is repulsive
and becomes attractive only near the critical point of the system
due to the critical fluctuation of the fluid density.}
\label{the_force_T}
\end{figure*}

We recall that for $L \gg L_{\rm crit}$, $\omega_s>0$, and
$\omega_l>0$, in line with Eq. (\ref{omega}) the behavior of the
Casimir force can be decomposed according to Eq. (\ref{manyXf}). The
aim of the current study is to determine  the leading finite-size
behavior of the Casimir force in any region of the thermodynamic
parameter space of the model. As stated in Sec. \ref{intro}, for any
$y$, $X^{{\rm sr}}(x,y)$ decays exponentially for
$|x|\rightarrow\infty$, while in that regime the other scaling
functions $X_{l}^{{\rm lr}}$ and $X_{s}^{{\rm lr}}$ decay according
to a power law. Based on the results of Refs.
\cite{DR2000,D2001,CD2002,DKD2003} one expects that the deviation
from the leading short-ranged behavior sets for $L\gtrsim \xi
\ln(\xi/\xi_0)$. For the scaling function $X_{l}^{{\rm lr}}$ one
expects $X_{l}^{{\rm lr}}(x\rightarrow\infty,0) = X_1 x^{\eta-2}$,
where $X_1$ is a universal constant.

Some information is also known about the asymptotic behavior of
$X_{s}^{{\rm lr}}(x\rightarrow\infty,0)$ \cite{MDB}. In Ref.
\cite{MDB} the authors have considered the behavior of the solvation
force in a fluid with short-ranged forces (i.e., $J^l=0$) under the
influence of long-ranged substrate potentials decaying at large
distances $z$ as $\delta v_s z^{-\sigma}$. In the case
$A_l=A_s=A_{l,s}=0$ (i.e., for $T>T_c$ in a magnetic system in the
absence of an external bulk magnetic field) they found numerically
together with analytical mean-field arguments that the leading size
dependence of the force $f$ is of the order of $L^{-(\sigma+1)}$. If
valid this would lead to $X_{s}^{{\rm lr}}(x\rightarrow\infty,0)=
X_s x^{d/2-2+\eta/2}$. For a mean-field model with $d=4$ and
$\eta=0$ this would imply $X_{s}^{{\rm lr}}(x\rightarrow\infty,0)=
X_s$, i.e., this scaling function would tend to a constant away from
the critical point (or vary $\sim \ln x$).
Moreover, consistency
with the structure of Eq. (\ref{manyXf}) imposes the temperature
dependence $\sim t^{\beta-\nu}$ of the term $\propto
L^{-(\sigma+1)}$. This differs from the result in Ref. \cite{MDB},
where the authors analyzed the system away from the scaling regime
and purportedly reported that the corresponding term varies $\propto
t^{-1}L^{-(\sigma+1)}$.

Finally, we finish this section by recalling that if $L\leq L_{\rm
crit}$ (or if $\omega_s<0$ or $\omega_l<0$) an expansion of the type
given in Eq. (\ref{manyXf}) is not possible. In that case the
effects of the van der Waals interactions will be important
everywhere in the critical region of the system, even at the bulk
critical point.

\section{Finite-size behavior of film
free energies and of the Casimir force} \label{results}

In this section we investigate the finite-size behavior of the total
Casimir force within the full temperature range. Experimentally this
force is accessible as the so-called solvation force. In Subsec.
\ref{critical} we focus on the critical behavior; the finite-size
behavior off criticality is discussed in Subsec. \ref{regular}.
Before passing to a detailed analysis of the influence of the
different parameters on the behavior of the force, we first provide
a view on the  {\it typical} behavior of the total force
($s=\lambda=1$) as a function of $L/\xi_t$ at $\Delta \mu=0$ (i.e.,
at bulk coexistence) (see Fig. \ref{the_force_T}(a)), and as a
function of $L/\xi_\mu$ at the critical temperature $T=T_c$ (see
Fig. \ref{the_force_T}(b)). The results are obtained within the
model presented in the previous section choosing a film thickness
$L=50$ layers. (Compare Fig. \ref{the_total_force_normalized_T} for
$L=500$; however, here the forces are not normalized by their value
at $T_c$.) One observes that the force (neglecting the contributions
from the direct substrate-substrate interaction so that the
conclusion stated after Eq. (\ref{thesum}) does not hold for $f_{\rm
total}$ considered here) is always {\it repulsive} outside the
critical region while within the critical region it becomes {\it
attractive} due to the critical fluctuations of the density. One
observes in both cases for $L/\xi >30$ (where $\xi$ is either
$\xi_t$ or $\xi_\mu$)  that the contribution
$\beta(\sigma-1)(A_l+A_{l,s})$ (indicated as ``asymptote'') provides
a good approximation of the total force. Using Eqs. (\ref{at}) and
(\ref{ah}), and $\rho_c=1/2$, it can be shown that
\begin{equation}\label{AlpAls}
\beta(\sigma-1)(A_l+A_{l,s})=4s \rho_b
+\frac{4\pi^{\frac{d-1}{2}}}{\sigma}
\frac{\Gamma\left[\frac{1+\sigma}{2}\right]}
{\Gamma\left[\frac{d+\sigma}{2}\right]} \lambda K \rho_b (1-\rho_b),
\end{equation}
where $K=\beta J^l_{\rm sr}$. Note that, if $s>0$ and $\lambda>0$,
$\beta(\sigma-1)(A_l+A_{l,s})>0$, i.e., the force away from the
critical region is {\it repulsive}.

In the following subsection we analyze in detail the behavior of the
force within the critical region and study its dependence on $s$,
$\lambda$, and $L$.

\subsection{Critical behavior}
\label{critical}

Within the present mean-field approach, one has $\beta=\nu=1/2$,
$\eta=0$, and $\gamma=1$. Apart from logarithmic corrections, these
mean-field values hold for $d=d_c=4$, provided  the interactions
governing the system are not too long-ranged.  Since for genuine van
der Waals interactions $d=\sigma$, in the following we adopt
$d=\sigma=4$, which leads to:
\begin{eqnarray}
\label{hfreeenergyexpandMFs} f(t,\mu,L)& \simeq & L^{-4}\Big\{X
[L/\xi_t,(\beta \Delta\mu) (L/\xi_{0,\mu})^{3},  \\ && l
\:(L/\xi_0)^{-2}, s \:(L/\xi_0)^{-1}] \nonumber \\ && + 3\beta
\left[A_l(T,\mu) + A_{l,s}(T,\mu) +A_s(T)\right] \Big\}.\nonumber
\end{eqnarray}
Note that for $d=\sigma=4$, $\eta=0$ the requirements $\sigma
>2-\eta$ and $\sigma>(d+2-\eta)/2$  for the irrelevance of the
long-ranged van der Waals type contribution to the behavior of the
force are fulfilled \cite{rem_Pavel}. For the model considered here
a direct estimate of $L_{\rm crit}$ yields $L_{\rm crit}=32 s$
(taking into account that here all distances are measured in units
of the lattice spacing). Since $s=1$ is the largest value of $s$ for
which we shall provide numerical results, we conclude that for
systems with $L\gg 32$ layers the van der Waals interaction is
expected to give only corrections to scaling while for smaller $L$
its effects will be important everywhere including the critical
region of the system and will affect even the determination of the
Casimir amplitude. Of course the above estimate does not tell the
exact meaning of ``much larger than'' 32, but we expect that a
factor of $10$ should put one onto the safe side. Thus the
expectations is that for $L\gtrsim 300$ the van der Waals
interaction will provide only corrections to scaling within the
critical region. Our subsequent analysis will demonstrate that this
expectation is indeed valid. We shall consider films with
thicknesses $L=50, 100, 500$, and $L=1000$ layers. It will turn out
that  for $L=500$ and $L=1000$ the van der Waals effects are small,
whereas for $L=50$ and $L=100$ it will turn out that they affect the
behavior of the force for all values of the thermodynamic
parameters.

In order to determine the scaling function $X$ of the Casimir force
one has to solve numerically Eq. (\ref{eqstatestandard}) for
$\phi^*({\bf r})$ and to insert this solution into Eq.
(\ref{freeenergyfinal}), where $\rho^*=(1+\phi^*)/2$.   Due to the
symmetry of the system one has $\phi^*({\bf r})=\phi^*(z)$. We
determine the force from the lattice version of Eq. (\ref{def}) (see
Eqs. (\ref{freeenergyfunctionalstarting}) and (\ref{omega})):
\begin{eqnarray}
f(L,t|l, s=0) &=& -\frac{\beta}{2}\left[\omega_{ex}(L+1,t|l,
s=0)\right.\nonumber \\ && \left. - \omega_{ex}(L-1,t|l,
s=0)\right]\;. \label{latforce}
\end{eqnarray}
(There are alternative approximations of $\partial \omega_{ex}/\partial L$ which, however, yield the same leading behavior in $L$ we are interested in.)

According to Appendix \ref{dereqstate}, for $d=\sigma=4$ the
equation for the order parameter reads (see Eq. (\ref{d4em})):
\begin{widetext}
\begin{equation}
\label{d4ems} {\rm arctanh}\left[ \phi^*(z)\right]= \frac{1}{2}\beta
\left[\Delta \mu-\Delta V(z)\right]+ K \left\{a_4 \, \phi^*(z) +
a_4^{nn} \left[\phi^*(z+1)+\phi^*(z-1)\right]+\lambda \sum_{z'=0
\atop |z'- z|\ge 2}^{L}g_4(|z-z'|)\phi^*(z')\right\},
\end{equation}
where $K=\beta J^l_{\rm sr}$, $\lambda=J^l/J^l_{\rm sr}$ (Eq.
(\ref{lambda})), and (see Eq. (\ref{g4ap}))
\begin{eqnarray}
g_4(a)&=& \pi^{3/2}\int_0^\infty dt\
t^{3/2}E_{4,4}(-t^4)\exp\left[-t
a^2 \right] \nonumber \\
&= & \frac{\pi^2}{2^{3/4}}\left \{   \left[
1+\left(\sqrt{2}a^2+1\right)^2
\right]^{1/4}\left[\sin\left(\frac{1}{2}{\rm arccot}[\sqrt{2}a^2+1]
\right) -\cos\left(\frac{1}{2}{\rm
arccot}[\sqrt{2}a^2+1] \right)\right] \right. \nonumber \\
& & +\left. \left[ 1+\left(\sqrt{2}a^2-1\right)^2
\right]^{1/4}\left[\sin\left(\frac{1}{2}{\rm arccot}[\sqrt{2}a^2-1]
\right) +\cos\left(\frac{1}{2}{\rm arccot}[\sqrt{2}a^2-1]
\right)\right]\right \}. \label{g4}
\end{eqnarray}
\end{widetext}
In Eq. (\ref{d4ems}) $a_4=7+\lambda (c_4-4)$, $a_4^{nn}=1+\lambda
(c_4^{nn}-1/2)$, where $c_4=4.900$ and $c_4^{nn}=1.028$ are
constants evaluated in Eqs. (\ref{c2}) and (\ref{c2nn}),
respectively. It is straightforward to show that the critical
coupling of the {\it bulk} system is
\begin{equation}\label{Kc}
K_c^{-1}(\lambda)=a_4(\lambda)+2a_4^{nn}(\lambda)+2\lambda
\sum_{z\ge 2} g_4(z),
\end{equation}
so that the bulk critical point coordinates are
$(K=K_c(\lambda),\Delta \mu=0)$. Note that the position of the
critical point depends on $\lambda$.

For $(+,+)$ boundary conditions one has $\rho(0)=\rho(L)\equiv 1$,
or equivalently $\phi(0)=\phi(L)\equiv 1$. We have imposed the
boundary conditions in such a way that  the number of liquid layers
with independent degrees of freedom is $L-1$. (We recall our remark
in the Introduction that the Casimir force hinges on the definition
of what is called the film thickness.) The function $g_4(a)$
accounts for the fluid-fluid long-ranged interaction. We note that
in many analyses $g_4(a)$ is typically neglected;  even in numerical
treatments  of van der Waals systems the range of the interaction
between the fluid particles is often actually truncated rendering
effectively a short-ranged potential. One can cast Eq. (\ref{d4ems})
also into a continuum form which is a close analogue of the one
usually derived from a Ginzburg-Landau type Hamiltonian. The
corresponding generalized Euler-Lagrange equation for the profile
then reads (see Eq. (\ref{eqcont})):
\begin{eqnarray}\label{eqconts}
&& \phi^*(z)+\frac{1}{3}(\phi^{*}(z))^3=\frac{1}{2}\beta (\Delta
\mu-\Delta V(z))\nonumber \\ && +K\Bigg\{a_4 \phi^*(z) + a_4^{nn}
\left[2 \phi^*(z)+\frac{d^2 \phi^*(z)}{dz^2}\right]\nonumber \\ &&
 +\int_{0}^{L}g_4(|z-z'|)\phi^*(z')dz' \Bigg\}.
\end{eqnarray}

We start our evaluation of the Casimir force by first reproducing
the analytically exactly available mean-field results for the force
for $(+,+)$ boundary conditions. That serves also as a valuable
check for the model we have described above.

For the following, Eq. (\ref{d4ems}) defines the model the
finite-size behavior of which will be investigated in detail. In
this model the long-ranged substrate fluid interaction is present
via the substrate potentials $\Delta V$, while the long-ranged
fluid-fluid interaction is reflected by $\lambda \ne 0$.
Accordingly, the fully short-ranged model corresponds then to
$\Delta V=0$ and $\lambda=0$.

Before starting with the study of the influence of the different
parts of the interaction on the behavior of the force we briefly
comment on the universality of the scaling function $X$ in Eq.
(\ref{hfreeenergyexpandMFs}) within mean-field theory.

\subsubsection{Modified finite-size scaling for mean-field systems}
\label{mod_fss}

Here we assume that $L$ is large enough so that one is allowed to
consider only the scaling fields which are relevant in
renormalization-group sense. In this case, within non-mean-field
theories  hyper-scaling and  hyper-universality are valid so that
the finite-size behavior of the singular part of the grand canonical
potential density $\beta \omega_s \equiv \beta \Omega_s/V$ near the
bulk critical point of the system $(T=T_c, \mu=\mu_c)$ is given by
\cite{PF84}
\begin{equation}\label{fss_standard}
\beta
\omega_s(T,\mu,L)=L^{-d}X_\omega\left(\frac{L}{\xi_t(\lambda)},
\frac{L}{\xi_\mu(\lambda)}\right),
\end{equation}
where $\xi_t \equiv \xi_\infty(T\to T_c^\pm, \mu=\mu_c)= \xi_0^\pm
(\lambda) |t|^{-\nu}$ and $\xi_\mu \equiv \xi_\infty(T= T_c,
\mu\to\mu_c)= \xi_{0, \mu}(\lambda) |\beta_c \Delta
\mu|^{-\nu/\Delta}$ are the second moment bulk correlation lengths.
Here, as before, all lengths are measured in units of the lattice
spacing $a$, $t=(T-T_c)/T_c$, $\Delta \mu = \mu-\mu_c$, and the
parameter $\lambda$  reflects the dependence of the bulk system on
the long-ranged component of the fluid-fluid interaction. Note that
there is no nonuniversal normalization factor in front of the
universal scaling function $X_{\omega}(x_t, x_\mu)$.  Indeed, due to
hyper-universality $\lim_{T\rightarrow T_c^+} \beta
\omega_s(T,\mu_c,\infty) \xi_t^d=Q$, where $Q$ is a universal
constant; this is consistent with the limit $L/\xi_t
\rightarrow\infty$ of Eq. (\ref{fss_standard}) with
$\lim_{x\rightarrow\infty} X_{\omega}(x,0)=Q\; x^d$.

Within mean-field theory hyper-universality is lacking and generates
a nonuniversal metric factor $A(\lambda)$, i.e.,
\begin{equation}\label{fss_mean_field}
\beta \omega_s(T,\mu,L)=L^{-4} A(\lambda) X_{\omega}^{\rm
MF}\left(\frac{L}{\xi_t(\lambda)},
\frac{L}{\xi_\mu(\lambda)}\right).
\end{equation}
It is easy to check that within our model the bulk grand canonical
potential density $\beta \omega_{s,{\rm bulk}}(T,\mu)$ of the system
is a universal function of $\beta/\beta_c$ and $\beta \Delta \mu$ in
the sense that $\beta \omega_{s,{\rm bulk}}(T,\mu)$ is not
proportional to any $\lambda$ dependent term. Requiring this
property to be compatible with Eq. (\ref{fss_mean_field}) in the
limit $L/\xi_t\to \infty$, it follows that $A(\lambda)$ has to be of
the form $A(\lambda)=Q_{\rm MF}\; [\xi_0^+(\lambda)]^4$, where
$Q_{\rm MF}$ is universal (i.e., independent of $\lambda$).
Furthermore, one has $X_{\omega}^{\rm
 MF}(x\rightarrow\infty,y)=x^4 \tilde{X}_{\omega}^{\rm MF}(x/y)$.
Moreover, since $\beta \omega_{s,{\rm bulk}}(T,\mu)$ does not depend
explicitly on $\lambda$ it follows that the ratio $\xi_t/\xi_\mu$
does not depend on $\lambda$, which for our model can indeed be
verified (see the following subsection). Thus Eq.
(\ref{fss_mean_field}) turns into the form
\begin{equation}\label{fss_mean_field_final}
\beta \omega_s(T,\mu,L)=\left(\frac{L}{\xi_0^+(\lambda)}\right)^{-4}
X_{\omega}^{\rm MF}\left(\frac{L}{\xi_t(\lambda)},
\frac{L}{\xi_\mu(\lambda)}\right),
\end{equation}
where $Q_{\rm MF}$ has been incorporated into the  scaling function
$X_{\omega}^{\rm MF}$. $X_{\omega}^{\rm MF}$ is ``universal'', in
the sense that it does not depend on $\lambda$ (neglecting the
contributions due to the irrelevant scaling fields).  Naturally, the
ratio $\beta \omega_s(T,\mu)/\beta_c \omega_s(T_c,\mu_c)$ defines a
scaling function that also does not depend  explicitly on $\lambda$ (see Eq.
(\ref{fss_mean_field})), but this ratio does not offer the
possibility to discuss the dependence of $\beta \omega_s(T,\mu)$  on
other parameters of the model near the critical point  because by
construction at the critical point this ratio equals $1$.

\subsubsection{The Casimir force in systems with
short-ranged interactions.}
\label{SRI}

\begin{figure*}[htb]
\begin{center}
\begin{tabular}{cc}
\hspace*{-.45cm}\resizebox{1.5\columnwidth}{!}
{\includegraphics{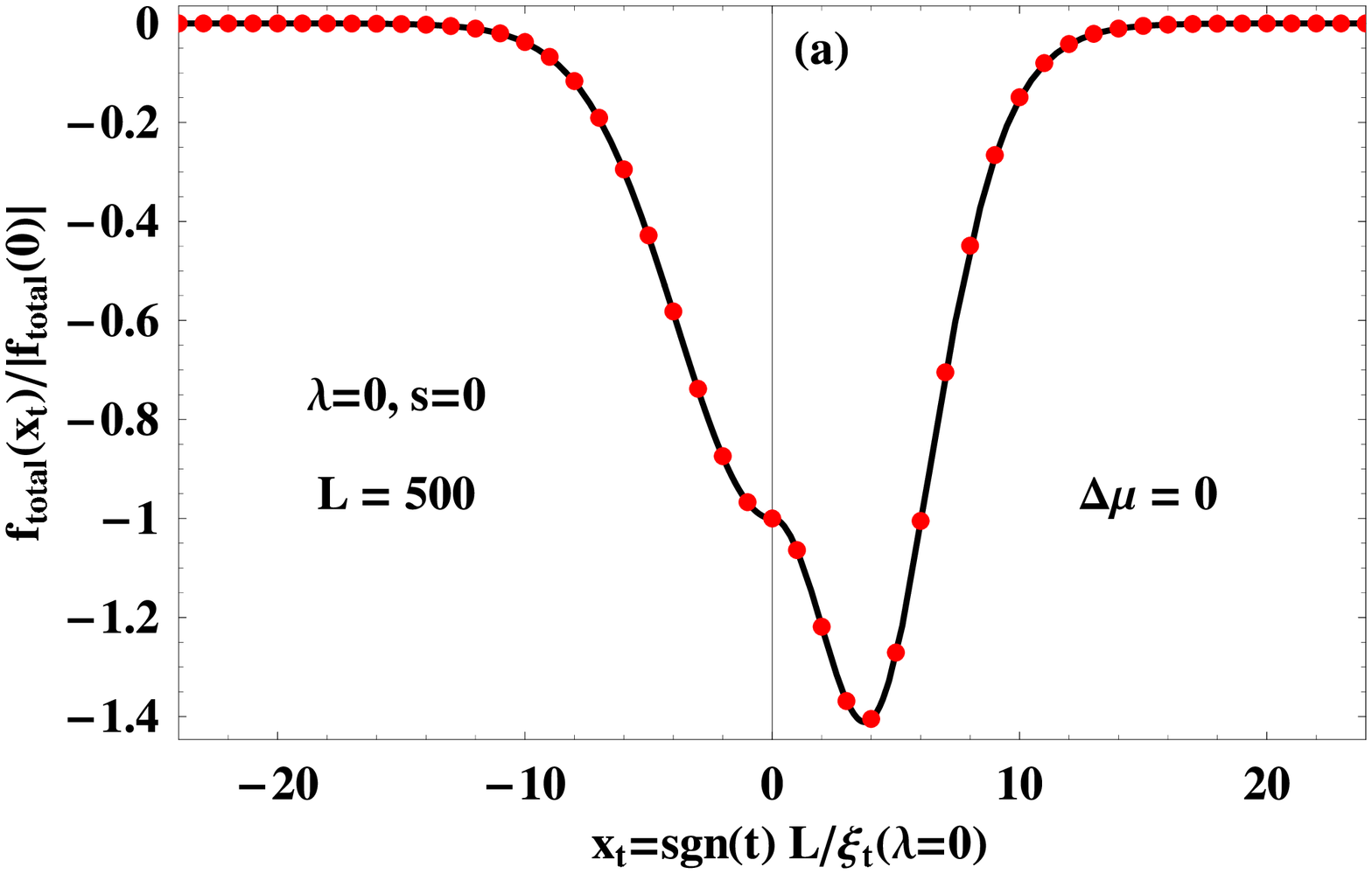}} &
\hspace*{-3.5cm}\resizebox{1.5\columnwidth}{!}
{\includegraphics{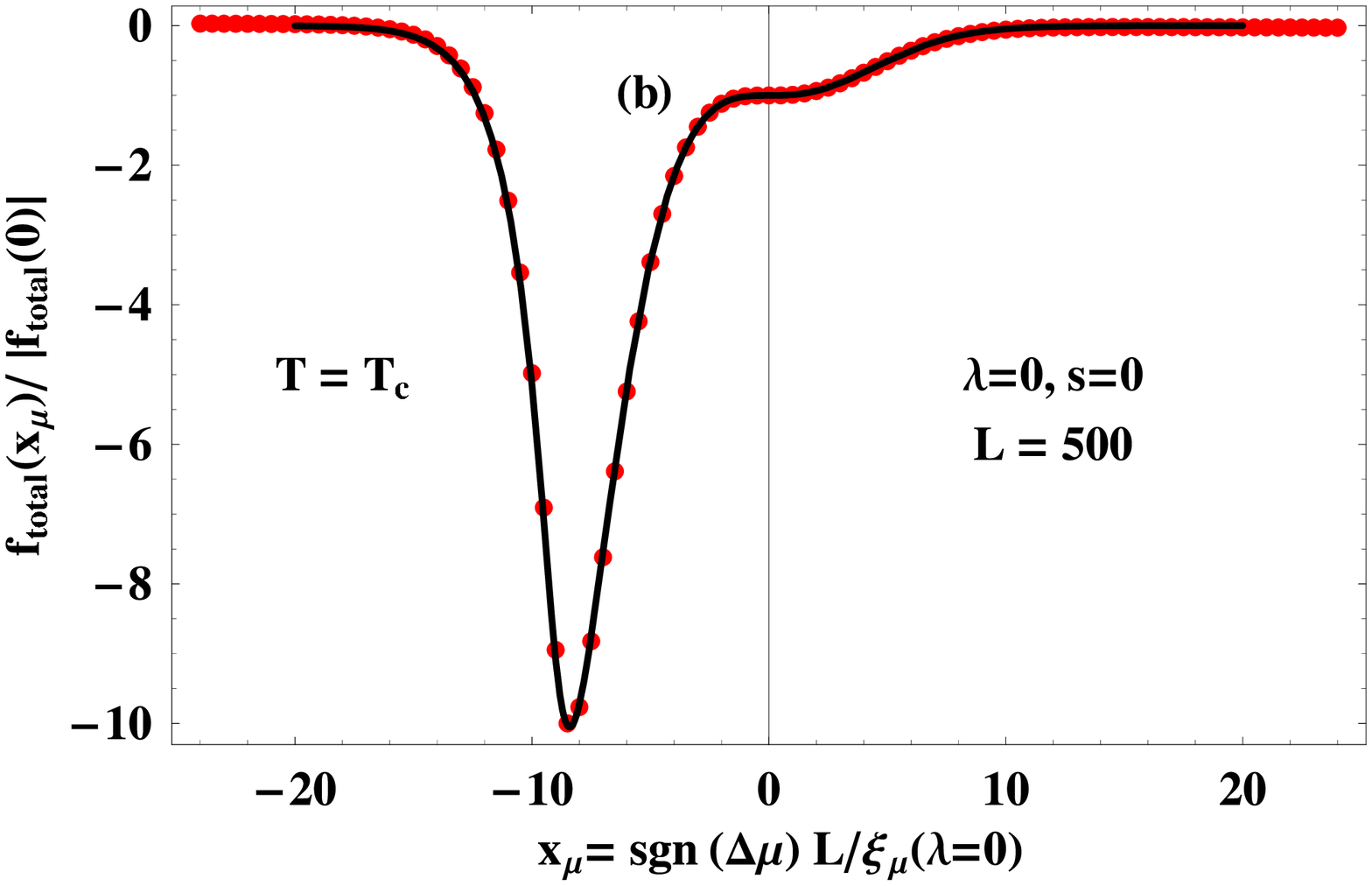}}
\end{tabular}
\end{center}
\caption{$(a)$ Comparison for $\Delta\mu=0$ between the behavior of
the {\it total} Casimir force in our mean-field {\it lattice} model
with short-ranged interaction (red dots) and the {\it singular} part
of the force obtained within {\it continuum} mean-field
Ginzburg-Landau theory \cite{M97} (black line). Both curves are
normalized by their absolute values at the bulk critical point. This
normalization is necessary because the Ginzburg-Landau theory
predicts the force only up to a nonuniversal multiplicative factor.
The lattice data have been calculated for  $L=500$. Both curves
overlap perfectly, which demonstrates that the corrections due to
the corrections to scaling, regular contributions, and the existing
arbitrariness in the definition of $L$ can be safely ignored if $L$
is of order the of $500$ and the interaction is short-ranged. Note
that the minimum occurs {\it above} the critical point at $x_t={\rm
sgn}(t) L/\xi_t \simeq 3.75$. $(b)$ The same as in $(a)$ for $T=T_c$
as a function of the field scaling variable $x_\mu=L/\xi_\mu$; the
results for the singular part of the force within a continuum mean
field Ginzburg-Landau type theory are from Ref. \cite{SHD03}. The
minimum of the force occurs at $x_\mu\simeq -8.4$, i.e., on the
vapor side of the bulk coexistence curve (with the walls preferring
the liquid phase). The force minimum is about a factor 10 deeper
than the force at $T_c$ and ca. seven times deeper than the minimum
along the temperature scan.} \label{comp_lattice_cont_short_ranged}
\end{figure*}
For $(+,+)$ boundary conditions and within mean-field theory the
analytical form of the scaling function of the Casimir force in
systems with short-ranged forces   has been obtained in Ref.
\cite{M97}. The corresponding result for $\Delta \mu =0$ is
\begin{subequations}\label{xsr}
\begin{equation}\label{xsr1}
 \text{(i)} \qquad X^{\rm sr}(y \ge 0)=-[2K(k)]^4
k^2(1-k^2),
\end{equation} with  $y=[2K(k)]^2(2k^2-1)$,
\begin{equation}\label{xsr2}\text{(ii)} \qquad
X^{\rm sr}(0\ge y \ge -\pi^2)=-4K^4(k),\end{equation}
 with  $y=[2K(k)]^2(2k^2-1)$,\begin{equation}\label{xsr3}
\text{(iii)}\qquad X^{\rm sr}(y\le-\pi^2)=-4K^4(k) (1-k^2)^2,
\end{equation} with  $y=-[2K(k)]^2(k^2+1)$,
\end{subequations} where $K(k)$ is the complete elliptic integral of
the first kind, $0\le k<1$. In Eq. (\ref{xsr}) the scaling variable
of $X^{\rm sr}(y)$ is $y=t\,L^{1/\nu}=t\,L^2$  (here $L$ is measured
in units of ${a}$, i.e.,  in the scaling variable $y$ $L$ is
dimensionless) and enters implicitly via $y=y(k)$ which can be
inverted uniquely to $k=k(y)$ because $y$ is a monotonic function of
$k$. We note that $X^{\rm sr}(y)$ is {\it analytic} for {\it all}
values of $y$, because the film critical point
$(T_c(L),\Delta\mu_c(L))$ is located off coexistence at
$\Delta\mu_c(L)\sim L^{-\Delta/\nu}\sim L^{-3}$ (see Fig.(\ref{pd}))
\cite{FisherNakanishi,Nakanishi-Fisher}. Obviously one has $y\ge 0$
if $k\ge 1/\sqrt{2}$ (with $k=1/\sqrt{2}$ corresponding to the bulk
critical point), $0\ge y \ge -\pi^2$ if $1/\sqrt{2} \ge k \ge 0$
(with $k=0$ corresponding to the actual film critical point), and
$y\le-\pi^2$ if $-1<k<0$ (negative $k$ describe the region below the
bulk critical point).
\begin{figure*}[htb]
\begin{center}
\begin{tabular}{cc}
\resizebox{1.5\columnwidth}{!}{\includegraphics{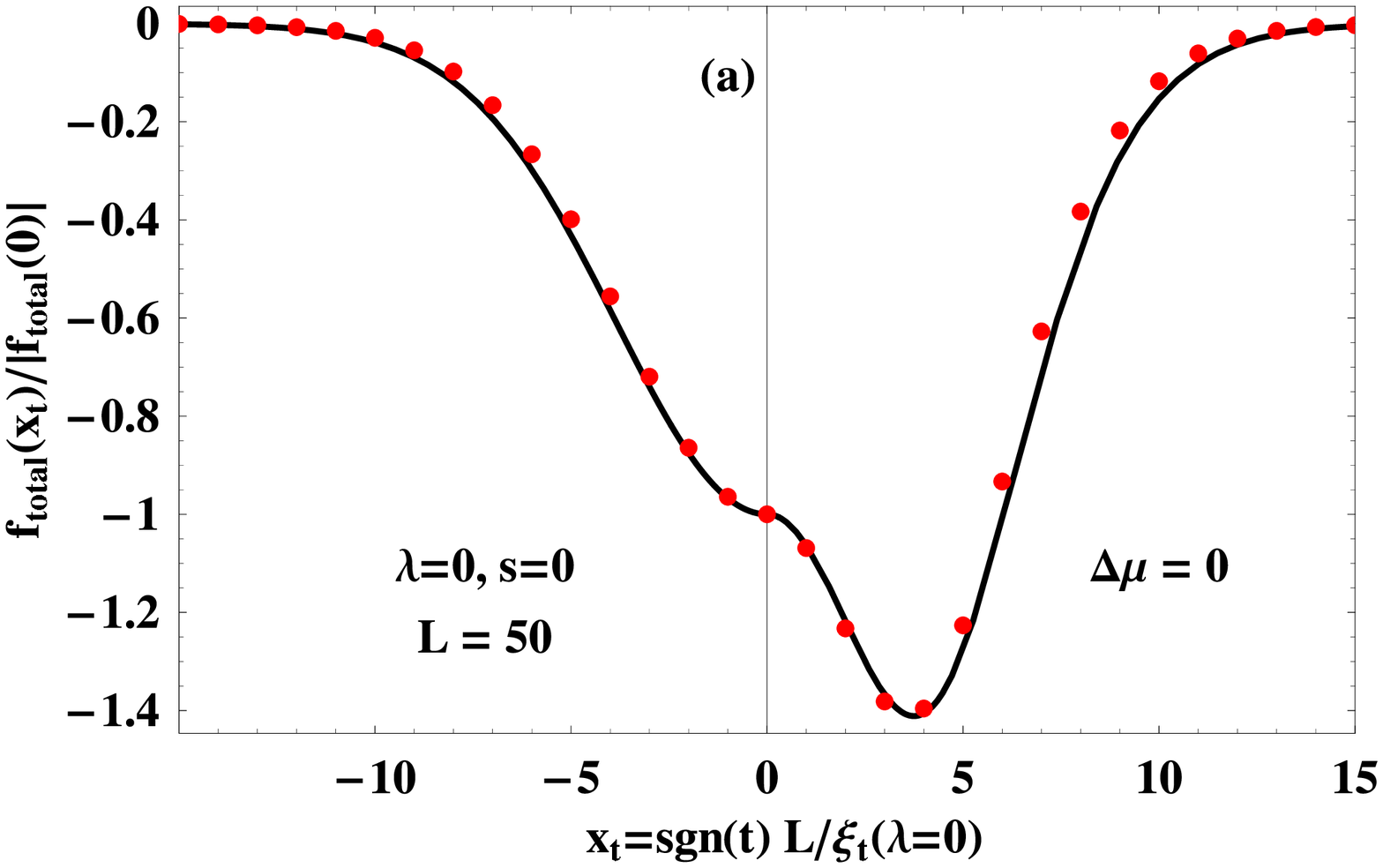}}
&
\hspace*{-3.75cm}
\resizebox{1.5\columnwidth}{!}{\includegraphics{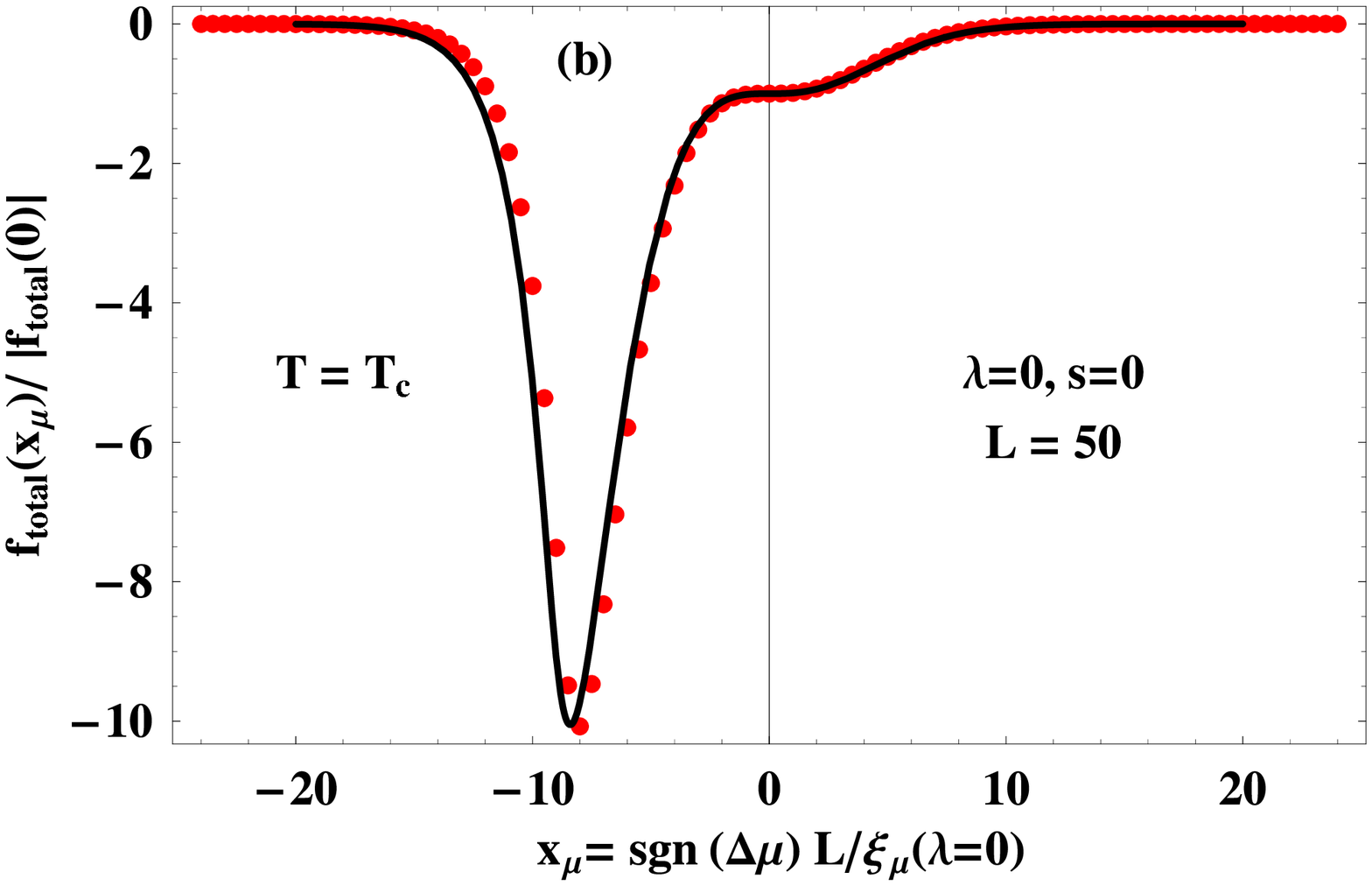}}
\end{tabular}
\end{center}
\caption{Comparison between the behavior of the {\it total} Casimir
force normalized by its absolute value at the bulk critical point
for a {\it lattice} model  with short-ranged interactions for $L=50$
(red dots) with the corresponding continuum Ginzburg-Landau result
(black curve).  $(a)$ For $\Delta\mu=0$ there are small differences
between the two  curves  for $|{\rm sgn}(t) L/\xi_t|\gtrsim 5$ where corrections
to scaling and the regular contributions start to show up. For
$|{\rm sgn}(t) L/\xi_t|\gg 1$ both curves decay exponentially and therefore they
cannot be distinguished for $|{\rm sgn}(t)L/\xi_t| \gtrsim 10$. $(b)$ Same as
$(a)$ for $T=T_c$ and as function of $\Delta \mu$ in terms of the
field scaling variable $x_\mu=L/\xi_\mu$. The deviations are largest
around the minimum.} \label{comp_L50}
\end{figure*}
\begin{figure*}[htb]
\hspace*{-0.28in}\includegraphics[angle=0,width=2.0\columnwidth]{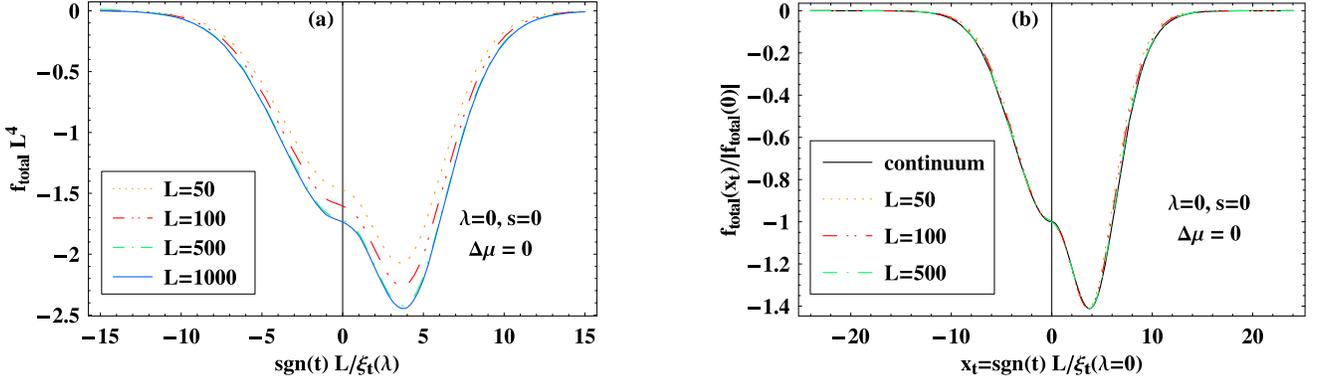}
\caption{Comparison between the behavior of the {\it total}
unnormalized $(a)$ and normalized $(b)$ Casimir force in a {\it
lattice} model with short-ranged interactions for different values
of $L$. While in $(a)$ the curves for $L=500$ and $L=1000$
practically coincide, which means that the nonuniversal corrections
are negligible, the corrections become apparent for  the film
thicknesses $L=50$ and $L=100$.  At the bulk critical point these
corrections amount to  $15\%$ for $L=50$ and $8\%$ for $L=100$.
$(b)$ If the curves from $(a)$ are normalized by their values at the
critical point they practically fall on top of  each other and
coincide with the continuum result.  This occurs because the force is
strongest at and near $T_c$ and, by construction, the curves for
different $L$ are forced to coincide at $T=T_c$ thus shifting the
possible deviations from each other towards large values of the scaling variable for which the force decays exponentially so that in that range  all curves again  coincide with each other. 
}
\label{comp_diff_L}
\end{figure*}
\begin{figure*}[htb]
\hspace*{-0.28in}\includegraphics[angle=0,width=2.0\columnwidth]{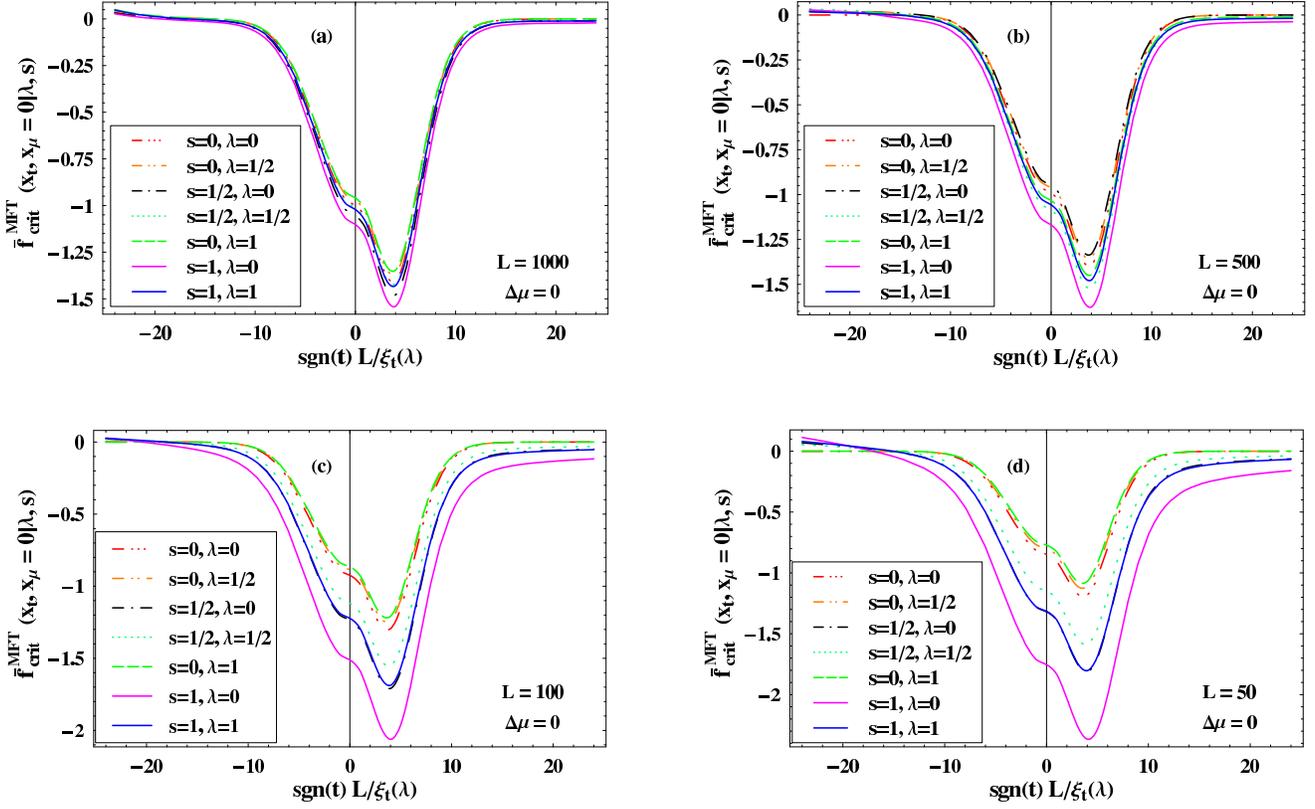}
\caption{Comparison between the behavior of the {\it critical}
Casimir force normalized according to Eq. (\ref{fcritdef_MF}) for
different values of $L$, $\lambda$, and $s$. The strength of the
long-ranged fluid-fluid interaction is reflected by the value of
$\lambda$ (the larger $\lambda$ the stronger is the long-ranged tail
of the interaction in comparison with the short-ranged part (Eq.
(\ref{lambda}))) and the strength of the long-ranged substrate-fluid
interaction is reflected by the value of $s$ (the larger $s$ the
stronger is the effective interaction with the substrate (Eqs.
(\ref{sdef}) and (\ref{s_def}))). We consider systems with $L=1000
\;(a), 500\; (b), 100\; (c)$, and $L=50\; (d)$ layers. If $L$
decreases one observes a more pronounced dependence on $s$ and
$\lambda$. While for $L=1000$ and $L=500$ all curves are very close
to each other, they significantly deviate from each other for
$L=100$ and $L=50$. For all $L$ the maximal deviations with respect
to the short-ranged case corresponding to $s=\lambda=0$ occur for
$s=1, \lambda=0$ (which strengthens the force) and $s=0,\lambda=1$
(which weakens the force). As nonuniversal  effects they  depend on
the absolute value of $L$. While for $L=1000$ the maximal deviation
is (still) about $10 \%$, for $L=50$ it reaches $100\%$.   This
profound dependence on the values of $s$ and $\lambda$ is most
visible for $L=50$. Contrary to the situation for $L=1000$ and
$L=500$, for $L=50$ none of the curves are  even close to each
other. The data show that the dependence on $s$ is more sensitive than that on $\lambda$:  changing the value of $s$ leads to more pronounced deviations from the results for short-ranged forces than changing the value of $\lambda$. Furthermore, for smaller $L$ the dependence on $s$ is stronger and the curves vanish less rapidly  for $x_t\to\infty$. We recall the asymmetry in the asymptotes of the total force for $x_t\to-\infty$ and $x_t\to \infty$ (see Fig. \ref{the_force_T}(a)). In order to obtain its critical contribution the asymptote for $T>T_c$ was subtracted. For this reason one observes for ${\bar{f}}^{\rm {MFT}}_{\rm{crit}}$  a tendency to increase for  $x_t\to -\infty$. \label{the_critical_force_all}}
\end{figure*}

In order to check universal aspects of our model and the reliability
of our numerical procedures,  for $\Delta \mu=0$ and $T$ close to
$T_c$ in Fig.\,\ref{comp_lattice_cont_short_ranged}(a) we compare
the results for $X^{sr}(x_t={\rm sgn}(t)L/\xi_t,x_\mu={\rm
sgn}(\Delta \mu)L/\xi_\mu) =0)$ obtained within the lattice model
(Eq. (\ref{gcpomega}) with $\Delta \mu=\Delta V=\Omega_{\rm
reg}=J^l=0$) with the above analytical results of the continuum
theory \cite{M97}. In addition, for $T=T_c$ and $\Delta \mu\ne 0$ in
Fig.\,\ref{comp_lattice_cont_short_ranged}(b) we present the scaling
function of the force $X^{sr}(x_t=0,x_\mu)$ as obtained within the
lattice model in comparison with the numerical results of the
continuum theory \cite{SHD03}.
Here $\xi_t$ is the bulk correlation length $\xi(t\rightarrow \pm
0,\Delta \mu=0)=\xi_0^\pm |t|^{-\nu}$, with $\nu=1/2$, while
$\xi_\mu(t=0,\Delta \mu \to 0)=\xi_{0,\mu}|\Delta
\mu|^{-\nu/\Delta}$ with $\nu/\Delta=1/3$. For the continuum model
$\xi_0^+=a$, $\xi_0^-=a/\sqrt{2} $ and $\xi_{0,\mu}=a/\sqrt[3]{3}$
\cite{SHD03,PHA91} (where $a$ has been introduced as a length scale
in order to achieve compatibility with the lattice model).

The Fourier transform $\tilde{g}({\bf k})$ of the density-density
correlation function
\begin{eqnarray}\label{grf}
g({\bf r}_1-{\bf r}_2)& \equiv & \langle \rho({\bf r}_1) \rho({\bf
r}_2) \rangle - \langle \rho({\bf r}_1) \rangle \langle \rho({\bf
r}_2)\rangle  \\ &=& \frac{1}{4}\left[ \langle \phi({\bf r}_1)
\phi({\bf r}_2) \rangle - \langle \phi({\bf r}_1) \rangle \langle
\phi({\bf r}_2)\rangle\right] \nonumber
\end{eqnarray}
has the form
\begin{equation}\label{gk}
\tilde{g}({\bf k})= \frac{1}{[\rho (1-\rho)]^{-1}+\beta
\tilde{w}({\bf k})}.
\end{equation}
From Eq. (\ref{grf}), using  $-4\tilde{J}({\bf k})=\tilde{w}({\bf
k})$ and the small $k$ expansion of $\tilde{J}({\bf k})$, one
obtains for the second moment correlation length $\xi$  (compare
with Eq. (6.9) in Ref. \cite{PHA91})
\begin{equation}\label{corlength}
\xi= \frac{\sqrt{v_2}}{\sqrt{\frac{\beta_c/\beta}{4 \rho
(1-\rho)}-1}}=\frac{\sqrt{v_2}}{\sqrt{\frac{T/T_c}{1-\phi^2}-1}}.
\end{equation}
Solving the bulk equation $\phi=\tanh [ \phi (\beta/\beta_c)+\beta
\Delta \mu/2]$ for the order parameter one obtains from Eq.
(\ref{corlength}) that in the lattice system under consideration
\begin{equation}\label{xi_der}
\xi_0^+ =\sqrt{v_2}, \qquad \xi_0^-=\xi_0^+ /\sqrt{2}, \qquad
\xi_{0,\mu}=\sqrt{v_2}/\sqrt[3]{3}.
\end{equation}
For a system with a fluid-fluid interaction given by Eq.
(\ref{Jfluid}) with $J^l=0$ (i.e., short-ranged interaction) and
discretized on a cubic lattice one has $v_2=1/(1+2d)$, so that for
$d=4$ one has  $\sqrt{v_2} \simeq 1/3$. Figures
\ref{comp_lattice_cont_short_ranged}(a) and
 \ref{comp_lattice_cont_short_ranged}(b) demonstrate that the
normalized scaling functions as obtained  within the continuum model
and  within the lattice model practically coincide once the
amplitudes $\xi_0$ and $\xi_{0,\mu}$ for the  two models are chosen
as described above. Thus on the mean-field level the present lattice
model is indeed suitable for investigating the Casimir force. We
emphasize that at $T=T_c$  the  minimum of the force does not occur
at bulk coexistence $\mu=\mu_c$ but at a nonzero value $\Delta \mu$,
determined for each $L$ via $x_\mu\equiv {\rm sgn}(\Delta \mu)
L/\xi_\mu\simeq -8.4$, i.e., on the gas side of the bulk coexistence
curve (with the walls preferring the liquid phase). Similarly, if
$\Delta \mu =0$ the minimum of the force is not at but above $T_c$,
given by $x_t={\rm sgn}(t) L/\xi_t \simeq 3.75$. We note that
Fig.\,\ref{comp_lattice_cont_short_ranged} actually provides a
comparison between the behavior of the {\it total} Casimir force in
a {\it lattice} model  with short-ranged interaction  with the {\it
singular} part of the force obtained within {\it continuum}
mean-field Ginzburg-Landau theory. If $L$ is small enough the
corrections to the universal behavior, which are due to the
corrections to scaling, to regular contributions,  and to the
existing arbitrariness in the definition of $L$ (see below as well
as Appendix B in Ref. \cite{DR2006}), will become visible. For
$L=50$ this is shown in Fig.~\ref{comp_L50}. The role of $L$ is
visualized in Fig. \ref{comp_diff_L}. Since the Casimir force is a
derivative of the excess free energy with respect to $L$ it depends
on the definition of $L$ for a given system. In the current analysis
this distance is taken to be the lattice spacing times the number of
layers with independent degrees of freedom (which is $L-1$) plus two
times half the distances between the outermost layers with frozen
degrees of freedom and their adjacent inner layers, i.e., $L a$.

\subsubsection{The Casimir force in systems with van der Waals type
interactions}

In this subsection we analyze the effect of the range of the
fluid-fluid and the substrate-fluid interactions on the behavior of
the Casimir force, i.e., we consider systems in which either
$\lambda$ or $s$, or both, are nonzero. Note that $\lambda \ne 0$
implies $l \ne 0$ (see Eq. (\ref{bdef})). To this end we first
determine the bulk correlation length amplitudes $\xi_0$ and
$\xi_{0,\mu}$ which depend on $\lambda$. Due to Eq. (\ref{xi_der})
this requires to calculate  the Fourier coefficient $v_2$, for which
we obtain
\begin{equation}\label{v2}
    v_2=\frac{1+ \frac{\lambda}{2d}
\sum_{{\bf r}}\frac{|{\bf r}|^2}{1+|{\bf r}|^{d+\sigma}}\;
\theta(|{\bf r}|-1)}{1+2d+\lambda \sum_{{\bf r}}\frac{1}{1+|{\bf
r}|^{d+\sigma}}\; \theta(|{\bf r}|-1)}.
\end{equation}
For $d=\sigma=4$ and for a hypercubic lattice a numerical evaluation
yields
\begin{equation}\label{v2det}
    v_2=\frac{1+0.829 \; \lambda}{9+2.152 \; \lambda}.
\end{equation}
For $\lambda=0,\; 1/2, \; 1$, and $2$ one thus obtains $\xi_0^+=
\sqrt{v_2}=\!\ 1/3, 0.375, 0.405$, and $0.447$, respectively.
\begin{figure*}[htb]
\hspace*{-0.28in}\includegraphics[angle=0,width=2.0\columnwidth]{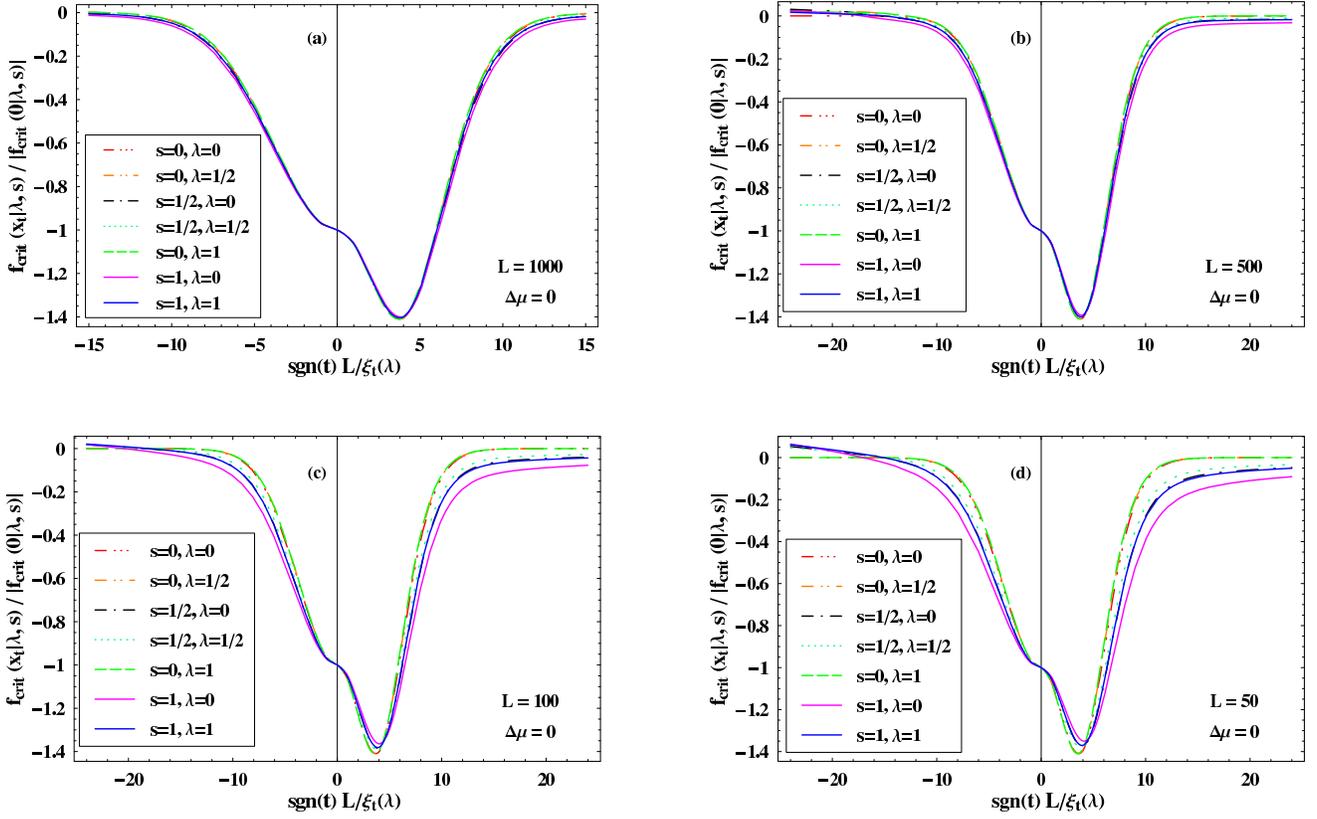}
\caption{Same data as in Fig. \ref{the_critical_force_all} but
normalized by the value of the critical Casimir force at the bulk
critical point. This normalization enforces coincidence of the
curves at $x_t=0$. The figure demonstrates that this kind of
normalization suppresses the visibility of the influence of the
long-ranged interactions, which is actually rather strong for $L=50$
and $L=100$ (see Fig. \ref{the_critical_force_all}). Nevertheless, for such values of $L$ and large scaling arguments the data demonstrate the dominant character of the finite-size contributions to the force which are due to the van der Waals fluid-fluid ($\lambda$) and substrate-fluid interactions ($s$).
\label{the_critical_force_normalized_all}}
\end{figure*}
\begin{figure*}[htb]
\hspace*{-0.28in}\includegraphics[angle=0,width=2.0\columnwidth]{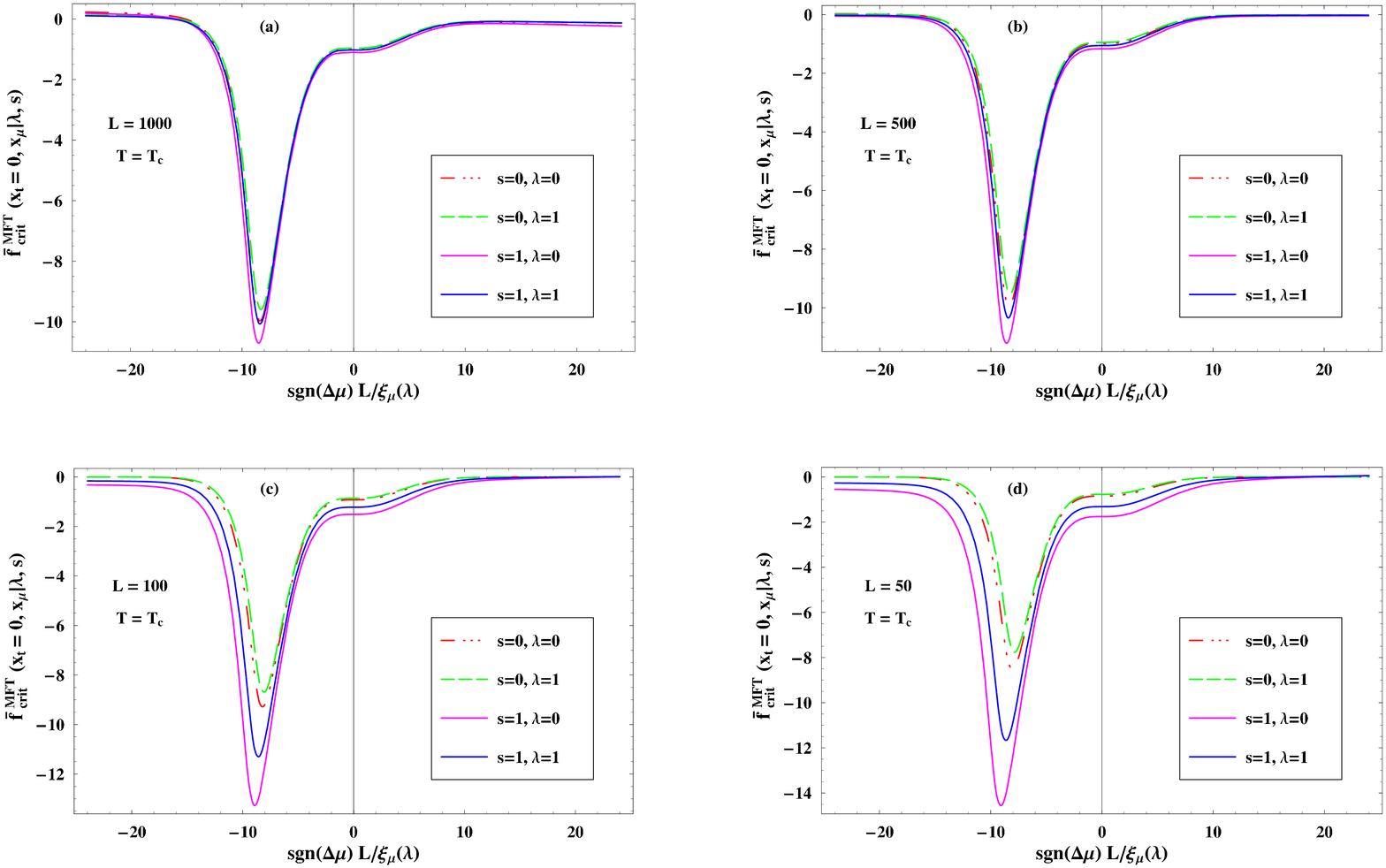}
\caption{Same as Fig. \ref{the_critical_force_all} as function of
the field scaling variable  $x_\mu={\rm sgn}(\Delta
\mu)L/\xi_\mu(\lambda)$ for $T=T_c$. The minimum of the force occurs
for {\it negative} $\Delta\mu$, i.e., on the gas side of the bulk
gas-liquid coexistence curve. The deepest minimum of the force
occurs for $s=1$, $\lambda=0$. For $L=1000$ the force at the minimum
is about 10 times stronger than the force at the critical point. If
$L$ decreases this minimum becomes even deeper; for $L=50$ the
corresponding force is about 15 times stronger than at the critical point. The force is weakest for
$s=0$, $\lambda=1$. Similar to Figs. \ref{the_critical_force_all} and \ref{the_critical_force_normalized_all} one again finds that the dependence on $s$ is more sensitive than that on $\lambda$:  changing the value of $s$ leads to more pronounced deviations from the results for short-ranged forces than changing the value of $\lambda$. We recall the asymmetry in the asymptotes of the total force for $x_t\to-\infty$ and $x_t\to \infty$ (see Fig. \ref{the_force_T}(b)). In order to obtain its  critical contribution the  asymptote for $T>T_c$ was subtracted.
\label{the_critical_force_H_all}}
\end{figure*}

\begin{figure*}[htb]
\hspace*{-0.28in}\includegraphics[angle=0,width=2.0\columnwidth]{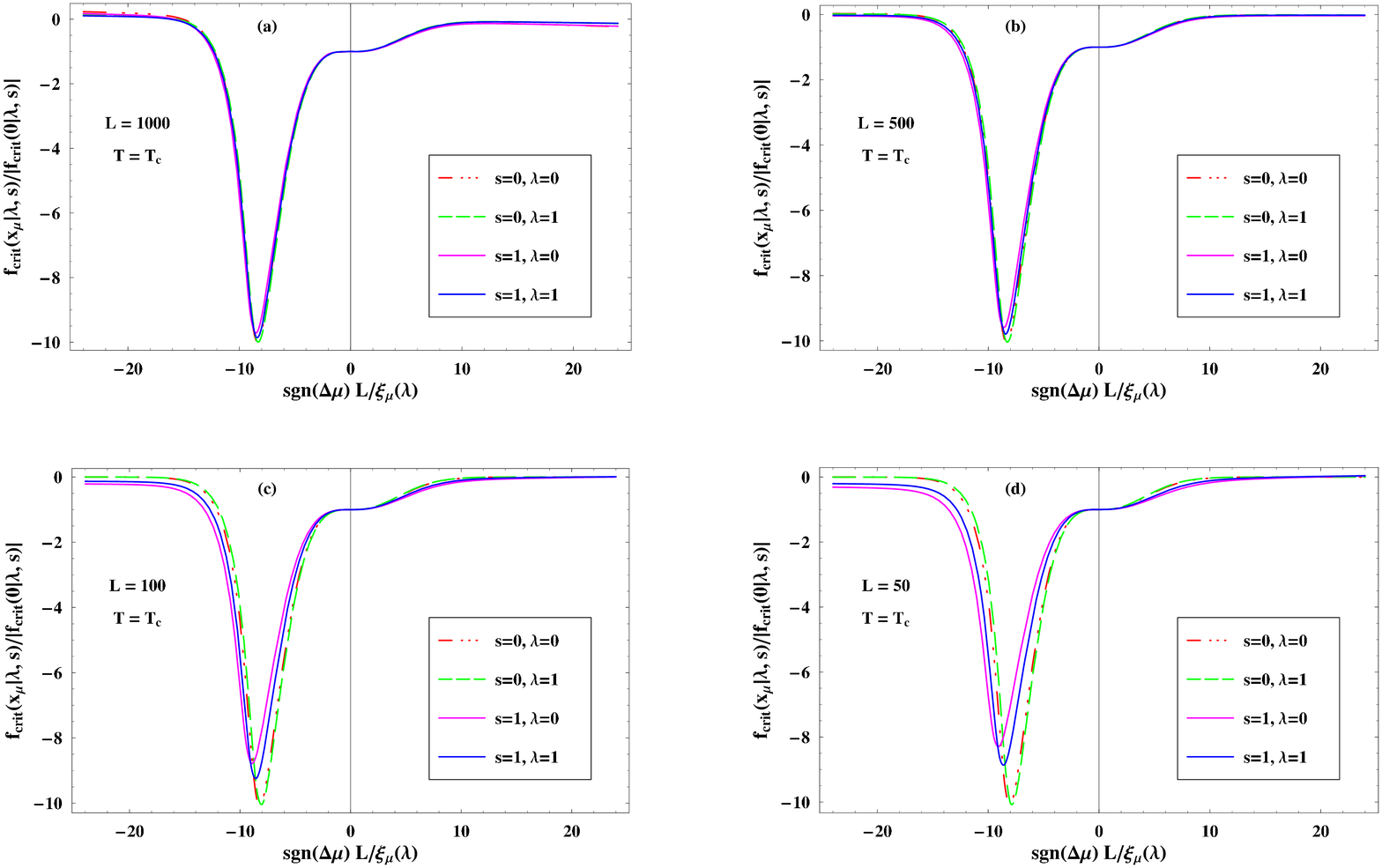}
\caption{Same data as in Fig. \ref{the_critical_force_H_all} but
normalized by the value of the critical Casimir force at the bulk
critical point. Whereas in Fig.
\ref{the_critical_force_normalized_all} as function of $x_t$ the
deviations occur for $x_t \gtrsim 10$, here the deviations are most
significant for $x_\mu \lesssim  -4$, affecting in particular the
minimum, but barely the scaling regime $x_\mu \gtrsim 0$. This is due to the fact that here the minimum of the force is rather deep (approximately $7$ times deeper than the corresponding one in Fig. \ref{the_critical_force_normalized_all}) so that the spread in the curves for $x_\mu \gtrsim 0$ is difficult to see, even for $L=50$, on the scale used in the plot.
\label{the_critical_force_normalized_H_all}}
\end{figure*}

The corresponding numerical results for the behavior of the critical
Casimir force $f_{\rm crit}$  are summarized in Figs.
\ref{the_critical_force_all},
\ref{the_critical_force_normalized_all},
\ref{the_critical_force_H_all}, and
\ref{the_critical_force_normalized_H_all}. ($f_{\rm crit}$ is that
part of the total force $f_{\rm total}$ calculated with the regular
background term subtracted; this procedure corresponds to the
analysis of experimental results if one subtracts from the value of
the force measured around $T_c$ the asymptote obtained by measuring
this force well {\it above} $T_c$.) While Figs.
\ref{the_critical_force_all} and
\ref{the_critical_force_normalized_all} deal with the temperature
dependence of the critical Casimir force at coexistence, Figs.
\ref{the_critical_force_H_all} and
\ref{the_critical_force_normalized_H_all} visualize the dependence
on the excess chemical potential at $T=T_c$. In Figs.
\ref{the_critical_force_normalized_all} and
\ref{the_critical_force_normalized_H_all} the force is normalized by
its value at the bulk critical point. This latter presentation does
not allow one to compare the forces for different $s$ and $\lambda$
at the critical point, because there by construction this ratio
equals $1$ for all parameters. This partial lack of insight from
this ratio can be overcome by choosing an alternative normalization,
which keeps the dependence on $\lambda$ and $s$ even at the critical
point:
\begin{eqnarray}\label{fcritdef}
&& \bar{f}_{\rm crit}(x_t,x_\mu|\lambda,s) \\ && =f_{\rm
crit}(x_t,x_\mu|\lambda,s)/|f_{\rm crit}(0,0|\lambda=0,s=0)|.
\nonumber
\end{eqnarray}
Here $f_{\rm crit}(0,0|\lambda=0,s=0)$ is the leading asymptotic
behavior $(d-1)\Delta_{+,+}L^{-d}$ of the critical Casimir force in
a system with short-ranged forces. This can be inferred from
experimental data of actual systems with dispersion forces by
considering the limit $L\gg L_{\rm crit}$ (Eq. (\ref{cL})) of the
difference between the total Casimir force $f_{\rm total}$ and the
regular background contribution (see Eq. (\ref{AlpAls}) and the
asymptote in Fig. \ref{the_force_T}) at the bulk critical point
$(T=T_c, \Delta\mu=0)$. In theoretical analyses $f_{\rm
crit}(0,0|\lambda=0,s=0)$ can be determined directly by switching
off the long-ranged parts of the interaction, i.e., by taking
$\lambda=0$ and $s=0$ from the outset. As explained in Subsec.
\ref{mod_fss}, within mean-field theory the rhs of Eq.
(\ref{fcritdef}) is not a ratio of {\it universal} scaling {\it
functions} (Eq. (\ref{fss_mean_field})) as it is the case for the
rhs of Eq. (\ref{fcritdef}) beyond mean-field theory, i.e., for
$d<4$. However, on the basis of the constructions in Subsec.
\ref{mod_fss} one can find a mean-field estimate for $\bar{f}_{\rm
crit}(x_t,x_\mu|\lambda,s)$ in $d=3$ by determining, e.g., within
our present mean-field approach,
\begin{eqnarray}\label{fcritdef_MF}
&& \bar{f}_{\rm crit}^{\rm MFT}(x_t,x_\mu|\lambda,s)\nonumber \\
&& =\Big\{\left[f_{\rm crit}(x_t,x_\mu|\lambda,s)/|f_{\rm
crit}(0,0|\lambda=0,s=0)|\right] \times \nonumber \\ &&
\left[\xi_0^+(\lambda=0)/\xi_0^+(\lambda)\right]^4\Big\}_{\rm MFT}.
\end{eqnarray}
Figs. \ref{the_critical_force_all} and
\ref{the_critical_force_H_all} show the temperature and field
dependence of $\bar{f}_{\rm crit}^{\rm MFT}$, respectively, for
$L=1000, 500, 100$, and $50$.

For decreasing values of $L$  one observes that the dependence on
$s$ and $\lambda$ becomes more pronounced. The maximum value of $s$
we have considered is $s=1$. As stated in the beginning of Subsect.
\ref{critical}, in this case one expects that for $L>300$ the van
der Waals interactions give rise only to corrections to the leading
short-ranged behavior and that for thinner films there are
significant deviations in the whole range of values of the scaling
variables. These expectations are confirmed by Figs.
\ref{the_critical_force_all} and \ref{the_critical_force_H_all}.
While for $L=1000$ and $L=500$ all curves are very close to each
other they deviate from each other significantly for $L=100$ and
$L=50$. Note that in all cases the force has a minimum {\it above}
the critical point.  The minimum is deepest for $s=1$, $\lambda=0$
and the force is weakest for $s=0$ and $\lambda=1$.

Figures \ref{the_critical_force_normalized_all} and
\ref{the_critical_force_normalized_H_all} show the same mean-field
data as in Figs. \ref{the_critical_force_all} and
\ref{the_critical_force_H_all}, respectively, but now normalized by
the actual values at the bulk critical point, i.e., $f_{\rm
crit}(x_t,x_\mu|\lambda,s)/|f_{\rm crit}(0,0|\lambda,s)|$. Whereas
the coincidence of the curves at $x_t=0$ and $x_\mu=0$,
respectively, is enforced by construction, for $L=50$ and $L=100$
there are significant differences for $|x_t|\gtrsim 10$, but barely
affecting the minimum at $x_t\simeq 3.75$ (Fig.
\ref{the_critical_force_normalized_all}), and for $x_\mu < -4$
including the minimum at $x_\mu=-8.4$, but barely affecting the
scaling regime $x_\mu \gtrsim 0$.
The occurrence of the force minimum at {\it negative} values of
$x_\mu$, i.e., on the gas side of the bulk gas-liquid coexistence
curve, can be understood by recalling that the Casimir force is a
fluctuation induced force. The fluctuations are  maximal at a small
negative bulk field which is needed to neutralize the ordering
effect (which suppresses the fluctuations) due to the
substrate-fluid potential and thus of the $(+,+)$ boundary
conditions. The ordering effect of the surface potentials can also
be neutralized by increasing the temperature to values slightly
above $T_c$. Therefore the magnitude of the force attains its
maximum for $T=T_c$ at a negative value of $x_\mu$ and for
$\Delta\mu=0$ at a positive value of $x_t$.

\subsection{Finite-size behavior off criticality: capillary
condensation regime}
\label{regular}

To a certain extent Fig. \ref{the_force_T} has already discussed
some features of the force outside the critical region, where the
total force is {\it repulsive}. Off criticality the force is well
approximated by $\beta (\sigma-1) (A_l+A_{l,s})$ given by Eq.
(\ref{AlpAls}). The small deviations $\Delta f$ of the force from
its asymptotes (see Fig. (\ref{the_force_T})) can be explained in
terms of an effective finite-size contribution $\Delta \mu_L$ to the
excess chemical potential $\Delta\mu$, which scales as $1/L$, i.e.,
 the confined fluid at $\Delta \mu$ has approximately the properties
of a bulk fluid at $\Delta \mu + \Delta \mu_L$ \cite{BE88}. This
implies ($\rho=(1+\phi)/2$) that $\phi \simeq \phi_b
(\Delta\mu+\Delta \mu_L)\simeq \phi_b(\Delta \mu)+\chi_b(\Delta \mu)
\Delta \mu_L$, with the second term producing a contribution $\Delta
f \sim 1/L$ relative to the leading first term.  We recall that the
total force discussed in Fig. \ref{the_force_T} has turned out to be
repulsive outside the critical region because it does not contain
the contribution $A_s$ from   the direct substrate-substrate
interaction. $A_s$ carries only a weak and smooth temperature
dependence via $\rho_s(T)$, which is not included in the degrees of
freedom considered  by the functional
(\ref{freeenergyfunctionalstarting}). If the contribution $A_s$ is
added to $f_{\rm total}$, the resulting force is {\it attractive}
(see Eq. (\ref{thesum})) throughout the whole thermodynamics space.
Besides the critical regime there is, however,   one additional
region  where the approximation of the force by $\beta (\sigma-1)
(A_l+A_{l,s})$ is also not valid and where the force  can be {\it
attractive}, too, even if $A_s$ is neglected.

\begin{figure}[h]
\includegraphics[angle=-90,width=\columnwidth,clip]{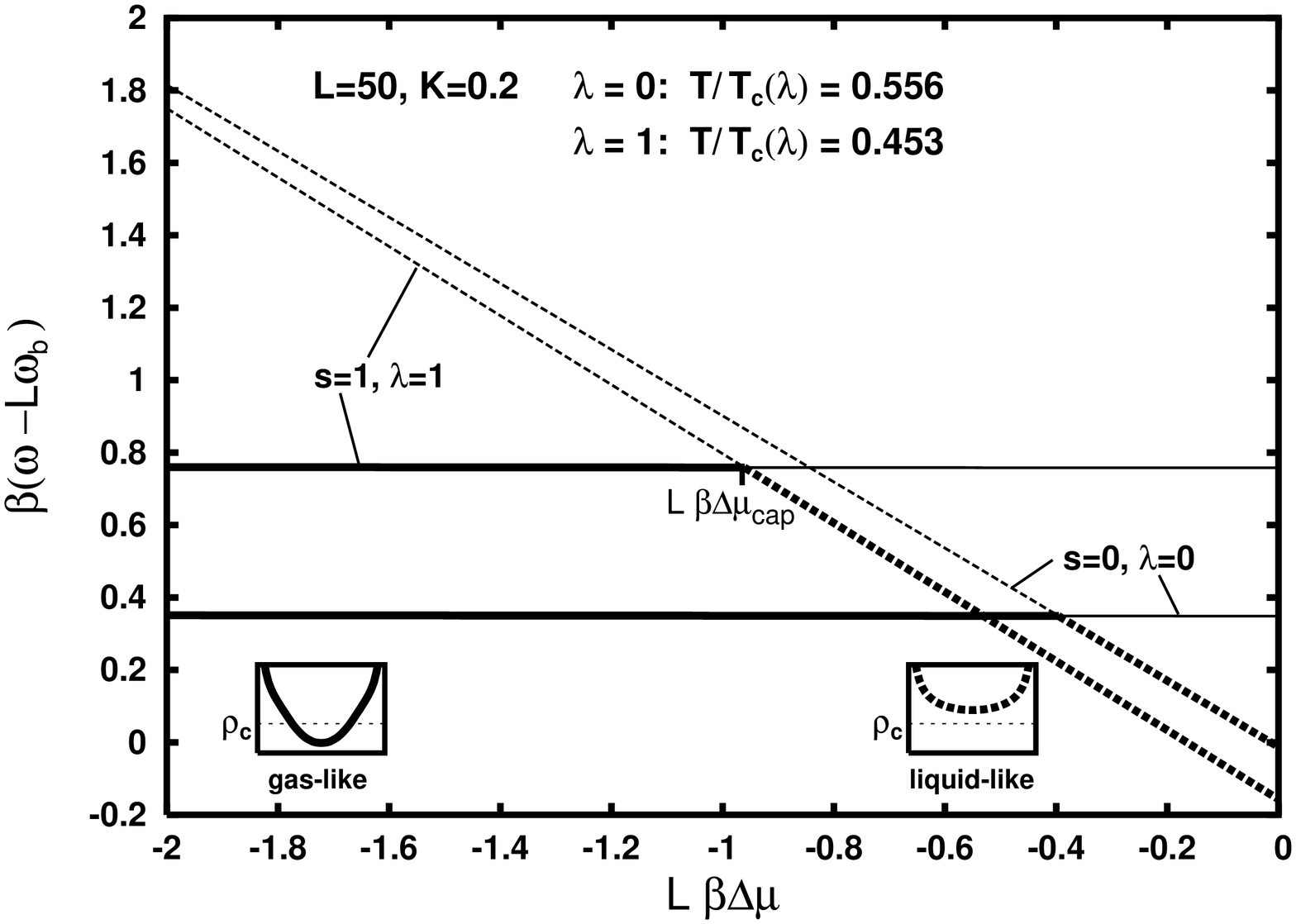}
\caption{Grand potential density in excess of the bulk contribution
as function of $L\,\beta \Delta\mu$ for the two competing profile
shapes (insets) in the capillary condensation regime for $(+,+)$
boundary conditions; $K=\beta J^l_{\rm sr}$. For $s=\lambda=0$ and
$s=\lambda=1$ the curves intersect at $L\,\beta \Delta\mu_{cap}
\approx -0.42$ and $-0.95$, respectively. The thermodynamically
stable states correspond to the minimum branches of $\beta \omega$. They are
presented by the thick parts of the lines. The thin parts of the lines indicate metastable states. The full lines correspond
to gas-like states, the short-dashed lines to liquid-like states.  \label{profile}}
\end{figure}

\begin{figure}[h]
\includegraphics[angle=-90,width=\columnwidth,clip]{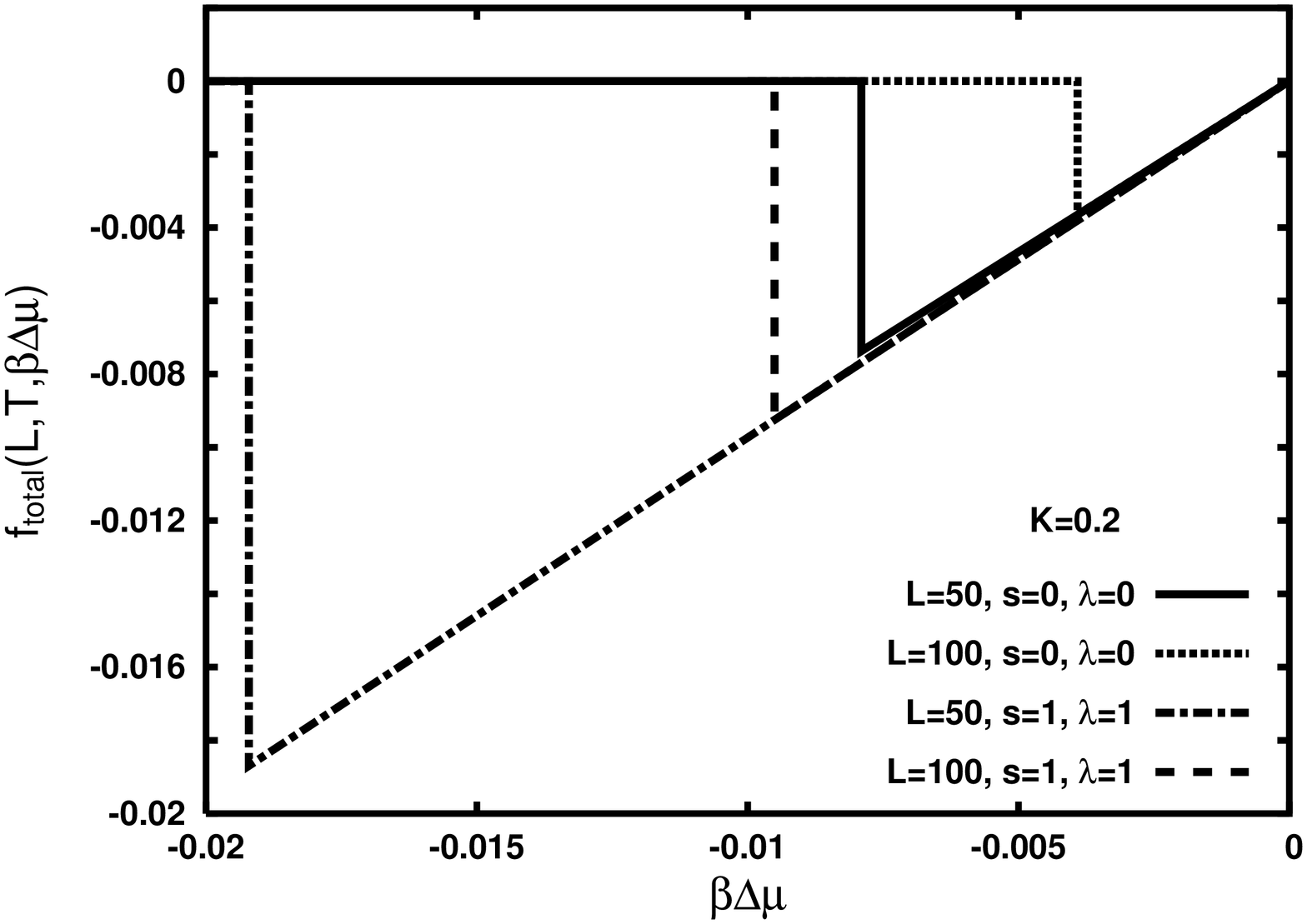}
\caption{Total force (obtained from Eq. (\ref{gcpomega})) per unit area and in units of $\beta$ as function of the chemical potential
$\Delta\mu<0$ for $\beta |\Delta \mu| L=O(1)$ and $T<T_c$.
The discontinuity of the force due
to capillary condensation occurs at $\Delta \mu_{\rm cap}(T,L)$. For $\Delta \mu
> \Delta \mu_{\rm cap}(T,L)$, with $\Delta
\mu_{\rm cap} \sim 1/L$, the leading behavior of the force is
basically independent of $L$ and is given by $2\phi_b\Delta \mu$. Note that the van der Waals contribution of the interaction strongly increases the maximal absolute value of the force. This occurs because both the long-ranged tails of the fluid-fluid ($\lambda$) as well as of the substrate-fluid ($s$) interactions promote the ordered state and increase the absolute value of $\Delta \mu_{\rm cap}$ beyond which the coexistence of a reservoir of gas with a liquid-like film is no longer possible. In this plot $K=0.2$ corresponds to  $T/T_c(\lambda)=0.556$ and $0.453$ for $\lambda=0$ and $\lambda=1$, respectively. Both for $\Delta \mu<\Delta \mu_{\rm cap}(T,L)$ and $\Delta \mu>0$ the leading behavior of the force is determined by the Hamaker term $\beta (\sigma-1) (A_l+A_{l,s})$ given by Eq. (\ref{AlpAls}). In these regions the magnitude of the force is of the order of $L^{-\sigma}$ (i.e., $L^{-4}$ within our model), while in the capillary condensed regime the maximal magnitude of the force  is of the order of $L^{-1}$. For this reason on the present scales the curves for $\Delta \mu< \Delta \mu_{\rm cap}(T,L)$ are not distinguishable.
\label{jump}}
\end{figure}

\begin{figure}[htb]
\includegraphics[angle=-90,width=\columnwidth]{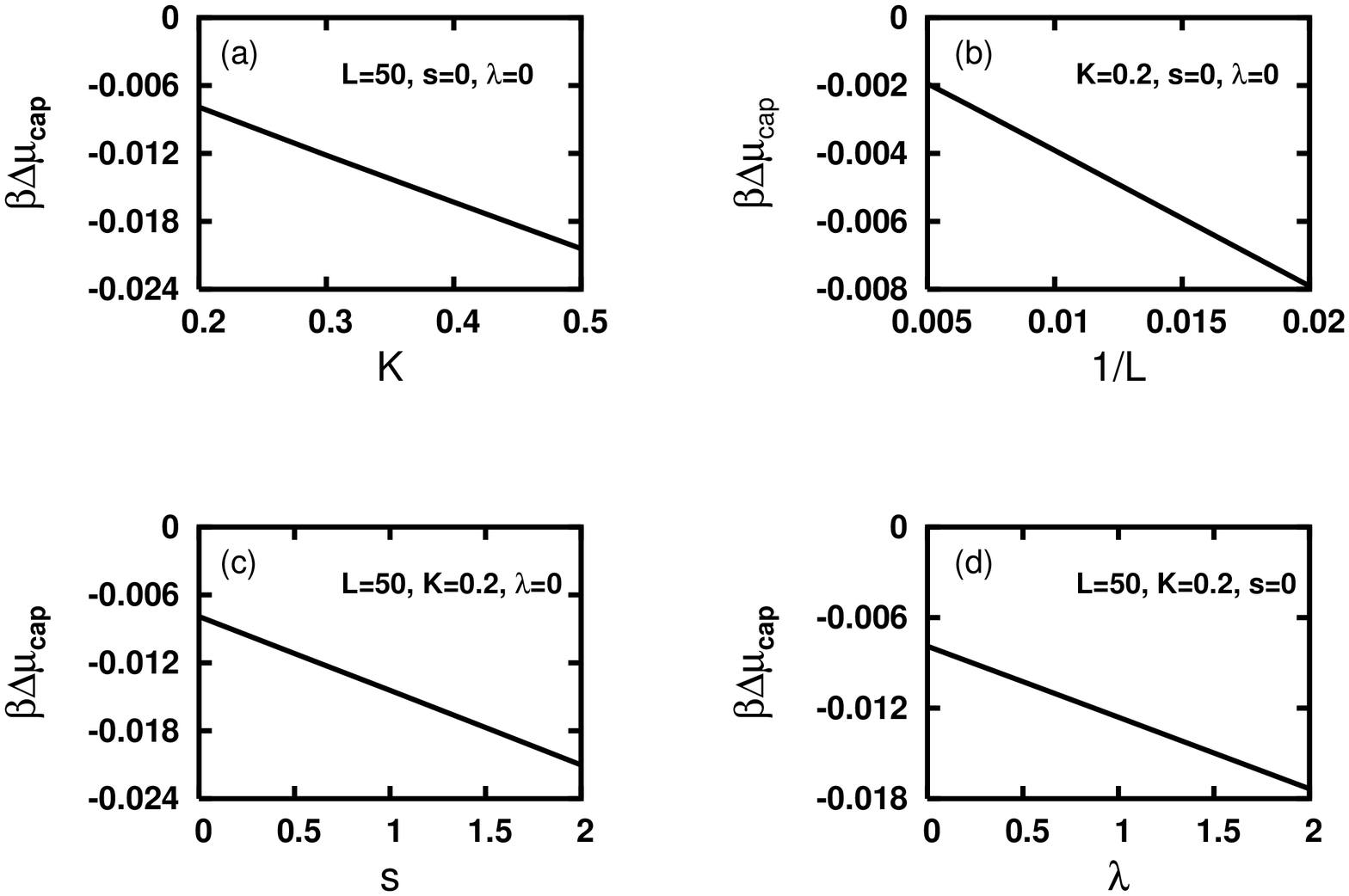}
\caption{ The curves illustrate the loci $\Delta \mu_{\rm
cap}(T,L|\lambda,s)$ for capillary condensation leading to a  jump
of the Casimir force as a function of the coupling $K=J_{\rm sr}^l/(k_B
T)$ $(a)$, the inverse linear size $L$ $(b)$, the relative strength
of the surface-fluid potential $s$ $(c)$, and the strength of the
van der Waals tail of the fluid-fluid interaction $\lambda$ $(d)$.
All these dependences turn out to be linearly decreasing within the
studied parameter range.  \label{parameters}}
\end{figure}

This is the so-called capillary condensation regime, which will be
discussed in the following.   This capillary condensation  regime
occurs at low temperatures $T<T_c$ and for $L \Delta \mu =O(1)$ with
$\Delta \mu <0$ if the walls prefer the liquid phase
\cite{SHD03,E90,EMT86,DME00}. In accordance with Fig. \ref{pd}
capillary condensation means that upon increasing the chemical
potential in a film of thickness $L$ there is a first-order phase
transition between spatially inhomogeneous gas-like and liquid-like
configurations at values $\Delta \mu_{\rm cap}<0$, i.e., before
reaching the bulk gas-liquid coexistence curve $\Delta \mu =0$. The
significant and interesting features of the force induced by
capillary condensation  are displayed in Figs.\,\ref{profile} and
\ref{jump}. For a film with $(+,+)$ boundary conditions, for $\Delta
\mu<0$ with $|\Delta \mu|$  sufficiently small it is favorable for
the system that $\phi \Delta \mu<0$, i.e., the equilibrium phase to
be liquid-like.  This means that $\phi$ is positive and thus follows
the  boundary conditions provided by the effective substrate
potentials. If $\Delta \mu$ becomes more negative, at a given
undersaturation  $\Delta \mu=\Delta \mu_{\rm cap}(T,L)$ the sign of
the order parameter changes abruptly to a negative value following
the bulk field and forcing the appearance of two interfaces within
the system. Accordingly, for large $L$ the corresponding
contribution in $\Omega$ which is linearly proportional to $\Delta
\mu$ changes by $2\phi_b \Delta \mu L$, thus leading to a
$L$-independent change $2\phi_b\Delta \mu$ in the force. (This
simple argument assumes that within this regime ($\Delta \mu
\lesssim 0$, low $T$) in a large part of the film the order
parameter profile $\phi(z)$  can be approximated by its bulk value
$\phi_b$.) Since the jump in the sign of $\phi$ occurs if the
advantage of having $\phi_b$ with the same sign as $\Delta \mu$,
i.e., $\phi_b \Delta \mu>0$ so that $\phi_b<0$, compensates the
disadvantage of creating two surface free energy contributions in
the system with a negative order parameter occurring inspite of
(+,+) boundary conditions, it is easy to understand that $\Delta
\mu_{\rm cap} \sim 1/L$.
In Fig.\,\ref{jump} the total force is shown as function of the
chemical potential $\Delta\mu$. For $\Delta \mu >\Delta\mu_{\rm
cap}$, up to a leading order, it does not depend on $L$. As
demonstrated numerically for large $L$ only the maximum absolute
value of the force is $L$-dependent, decreasing as $\sim 1/L$ due to
the $L$-dependence of $\Delta\mu_{\rm cap}(T,L)$.
Figures \ref{profile} and \ref{jump} present results for a system
with short-ranged interactions ($s=\lambda=0$) and a system in which
both the fluid-fluid and substrate-fluid interactions are
long-ranged ($s=\lambda=1$). It turns out that the behavior of the
force in the considered $(T, \mu)$ region depends on $s$ and
$\lambda$ mainly via the dependence of $\Delta \mu_{\rm cap}$ on $s$
and $\lambda$. This dependence is studied systematically in Fig.
\ref{parameters}. It shows that $\Delta \mu_{\rm cap}$ depends
linearly on $K$, $L^{-1}$, $s$ and $\lambda$. Furthermore the
maximal amplitude of the force also increases linearly if any of
these parameters increases.

\section{Summary and concluding remarks}
\label{summary}

In view of future experiments exploring the critical Casimir forces
in classical one- or two-component fluids confined by parallel
substrates with the same preference for their coexisting fluid
phases we have analyzed the actual and generic interplay between the
long-ranged fluctuation induced force and the long-ranged dispersion
forces acting both between fluid molecules and between fluid and
substrate molecules.

From general scaling arguments and explicit mean-field model
calculations we have obtained the following main results:
\begin{itemize}

\item[(1)]{Based on general renormalization group arguments
(Subsec.~\ref{scaling_dispersion}) we have established a
relevance-irrelevance criterion (Subsec.~\ref{criterion}) for the
importance of dispersion forces within the critical region of the
system. It turns out that if  the thickness $L$ of the liquid film
is much larger that $L_{\rm crit}$ (see Eq. (\ref{cL})), within the
critical region of the system the dispersion forces provide only
corrections to the leading behavior of the force. However, outside
the critical region, i.e., for temperatures $T \ne T_c$ or for
undersaturations $\Delta\mu\ne 0$, the influence of the dispersion
forces becomes important. This is of experimental importance because
it is difficult to thermodynamically position the system right at
the critical point.  In the opposite case, i.e., for $L \lesssim
L_{\rm crit}$ the contribution of the van der Waals (dispersion)
forces is always important and cannot be neglected even within the
critical region of the system, including  the bulk critical point.
Within a mean-field type model (Eqs. (\ref{omegaDF}),
(\ref{freeenergyfunctionalstarting}), (\ref{eqstatestandard}),
(\ref{Jfluid}), and (\ref{Jsubstrate})) we have studied the behavior
of the force as a function of the strength of the long-ranged part
of the fluid-fluid interaction, reflected by the parameter $\lambda$
(Eq. (\ref{lambda})), on the strength of the long-ranged part of the
effective substrate-fluid interaction, reflected by the parameter
$s$ (Eq.(\ref{s_def})), on the thickness $L$ of the film, and on the
(relevant) temperature and chemical potential scaling fields. The
interactions are taken to decay with the distance $r$ between the
fluid particles as $r^{-(d+\sigma)}$ and as $z^{-\sigma_s}$ for the
distance $z$ of a fluid particle from a single substrate. (Different
from other theoretical model calculations we have taken the
long-ranged tail of the fluid-fluid interaction fully into account,
which amounts to respecting the fact that the function $g_d(a)$ (Eq.
(\ref{g4}) and Fig. \ref{gplot}) is nonzero.) For a $d$-dimensional
system we consider the case $\sigma>2-\eta$ and
$\sigma_s>(d+2-\eta)/2$, which guarantees that the long-ranged tails
of the interactions  change neither the bulk nor the surface
universality classes \cite{PL83,D86}. The modified finite-size
scaling, which is needed if one -- as done here --  compares within
mean-field theory systems with different $\lambda$, is presented in
Eq. (\ref{fss_mean_field_final}). The results for the behavior of
the force as a function of $L/\xi_t$ and $L/\xi_\mu$, where $\xi_t$
and $\xi_\mu$ are the second-moment bulk correlation lengths for
$\Delta \mu=0$ and at $T_c$, respectively, are summarized in Figures
\ref{the_critical_force_all},
\ref{the_critical_force_normalized_all},
\ref{the_critical_force_H_all}, and
\ref{the_critical_force_normalized_H_all}. These explicit results
confirm  the relevance-irrelevance criterion mentioned above. The
numerical results presented  are for films with thicknesses
$L=1000$, $L=500$, $L=100$, and $L=50$ layers in our simple cubic
lattice model. We have varied $\lambda$ and $s$ between $0$
(short-ranged model) and  $1$. Figures \ref{the_critical_force_all},
\ref{the_critical_force_normalized_all},
\ref{the_critical_force_H_all}, and
\ref{the_critical_force_normalized_H_all} show that for $L=1000$ and
$L=500$ one observes universal scaling, i.e., independence  on the
values of $s$ and $\lambda$, whereas this is lost for  $L=100$ and
$L=50$, which is in line with $L\lesssim L_{\rm crit}$. The
comparison between Figs. \ref{the_critical_force_all} and
$\ref{the_critical_force_normalized_all}$ on one hand and Figs.
\ref{the_critical_force_H_all} and
\ref{the_critical_force_normalized_H_all} on the other hand shows
which kind of analysis of the effects enhances or suppresses the
visibility of the long-ranged forces. It turns out that $L_{\rm
crit}$ is surprisingly large (see Eq. (\ref{cL})) which enhances the
importance of taking into account the dispersion forces for the
interpretation of experimental data of critical Casimir forces in
fluid systems.}

\item[(2)]{If all long-ranged interactions are taken into account -
the fluid-fluid, the substrate-fluid, and the substrate-substrate
one, for the case of $(+,+)$ boundary conditions studied here we
confirm explicitly within our model that the effective interaction
between the substrates is attractive throughout the whole
temperature range (see Eq. (\ref{thesum})). The latter is described
by three different Hamaker constants which account for the
aforementioned interactions. If the direct substrate-substrate
interaction is omitted (which amount to subtracting the force from
the corresponding null experiment)  the force is repulsive (see
Figs.  \ref{the_total_force_normalized_T} and \ref{the_force_T})
except within the critical and the capillary condensation regime.}

\item[(3)]{For short-ranged systems, within the critical region our
lattice model reassuringly reproduces the universal scaling function
of the critical Casimir force as obtained from continuum field
theory, provided $L$ is sufficiently large (Subsec. \ref{SRI} and
Fig.\,\ref{comp_lattice_cont_short_ranged}). Surprisingly, even for
short ranged forces $(\lambda=s=0)$ $L=50$ and $L=100$ turn out to
be too small in order to reach the universal scaling function given
by the continuum field theory; $L=500$ and $L=1000$ are large enough
in this respect (see Figs. \ref{comp_L50} and \ref{comp_diff_L}).}

\item[(4)]{At low temperatures, $\Delta \mu <0$, and $L\Delta \mu =O(1)$
the force undergoes a jump as function of the excess chemical
potential $\Delta\mu$ at a certain value $\Delta \mu_{\rm cap}<0$
caused by capillary condensation (see Figs. \ref{pd},
\,\ref{profile}, and \ref{jump}). The position of the jump (Fig.
\ref{jump}) and its magnitude (Fig. \ref{profile}) vary linearly as
function of $1/T$, $1/L$, $s$, and $\lambda$ (Fig.
\ref{parameters}).}
\end{itemize}

\begin{acknowledgments}
D. Dantchev acknowledges the  hospitality of the
Max-Planck-Institute for Metals Research in Stuttgart and the
partial support via Project F-1402 of the Bulgarian Fund for
Scientific Research.
\end{acknowledgments}

\appendix

\section{Derivation of the effective Hamiltonian for a
van der Waals system confined between two parallel plates}
\label{EffectiveHamiltonian}

We consider a  fluid composed of particles interacting via a pair
potential $w^l({\bf r} - {\bf r}')=-4J^l({\bf r} - {\bf r}')$ and
bounded by substrates whose particles interact with the liquid
particles with a pair potential $w^{l,s}({\bf r} - {\bf
r}')=-4J^{l,s}({\bf r} - {\bf r}')$. Within the lattice gas model
for any given configuration ${\cal C}$ of particles
$\{p_i^s,p_j^l\}$, $i \in {\cal S}$, $j \in {\cal L}$,  where ${\cal
L}$ and ${\cal S}$ denote the spatial region in a simple cubic
lattice with lattice constant $a$ occupied by the liquid and
substrate particles, respectively, the energy of the fluid is given
by
\begin{eqnarray}
{\cal E}&=&\sum_{i\in {\cal S},j\in {\cal
L}}w_{i,j}^{l,s}p_{i}^{s}p_{j}^{l} + \frac{1}{2}\sum_{i,j\in {\cal
L}}w_{i,j}^{l}p_{i}^{l}p_{j}^{l}\nonumber \\
&=& -4\sum_{i\in {\cal S},j\in {\cal
L}}J_{i,j}^{l,s}p_{i}^{s}p_{j}^{l} -{2}\sum_{i,j\in {\cal
L}}J_{i,j}^{l}p_{i}^{l}p_{j}^{l};
\end{eqnarray}
$p_j^l\in \{0,1\}$ and $p_i^s \in \{0,1\}$ denote the occupation
numbers for the liquid and substrate  particles, respectively. Since
only the part $\{p_j^l\}$ corresponding to the fluid degrees of
freedom exhibit  criticality at $T_c$ and the fluctuations of the
substrate degrees of freedom $\{p_j^s\}$  are unimportant in this
respect, one can replace the latter ones by their mean-field values.
If the fluid is in contact with a particle reservoir at a given
(excess) chemical potential $\mu$ and temperature $T$, the partition
function for the liquid is
\begin{eqnarray}
\label{semiinfinite} Z &=&\sum_{\{p^l_j\}}\exp{\left[-\beta
\left({\cal E}-\mu \sum_{j \in {\cal L}} p_j^l\right)\right]}
\nonumber \\
&=& \sum_{\{p^l_j\}}\exp\Bigg[\beta \Bigg( 4\sum_{i\in {\cal S},j\in
{\cal L}}J_{i,j}^{l,s}\rho_{i}^{s}p_{j}^{l}\nonumber  \\
&& +{2} \sum_{i,j\in {\cal
L}}J_{i,j}^{l}p_{i}^{l}p_{j}^{l}+\mu\sum_{j \in {\cal L}}
p_j^l\Bigg)\Bigg],
\end{eqnarray}
where $\rho_i^s\equiv \langle p_i^s\rangle$. We assume that the
solid substrate is only weakly influenced by the presence of its
surface, so that $\rho_{i}^{s} =\rho_s={\rm const}$ in ${\cal S}$,
which leads to
\begin{eqnarray}
Z &=& \sum_{\{p^l_j\}}\exp\Bigg[\beta \sum_{j\in {\cal L}}
\Bigg(4\rho_s \sum_{i\in {\cal S}}J_{i,j}^{l,s}+\mu \Bigg)
p_{j}^{l}\nonumber \\ && + {2}\beta \sum_{i,j\in {\cal
L}}J_{i,j}^{l}p_{i}^{l}p_{j}^{l}\Bigg].
\end{eqnarray}
Modeling the pair potentials as
\begin{eqnarray}
J^l_{i,j} &\equiv & J^l_{\rm sr}\left\{\delta(|{\bf r}_i-{\bf
r}_j|)+\delta(|{\bf r}_i-{\bf r}_j|-1)\right\}\nonumber \\ &&
+J^l/(1+|{\bf r}_i-{\bf r}_j|  ^{d+\sigma})\; \theta(|{\bf r}_i-{\bf
r}_j|  -1),
\end{eqnarray}
(note that $\theta(0)=0$ and $\delta(x)=1$ for $x=0$ and zero
otherwise) and
\begin{eqnarray}
J^{l,s}_{i,j} &\equiv & J^{l,s}_{\rm sr}\,\delta(|{\bf r}_i-{\bf
r}_j|-1)\nonumber \\ && +J^{l,s}/|{\bf r}_i-{\bf r}_j|^{d+\sigma_s}
\; \theta(|{\bf r}_i-{\bf r}_j|-1)
\end{eqnarray}
one finds for $\sum_{i}J_{i,j}^{l,s}$:
\begin{widetext}
\begin{eqnarray}
\sum_{i}J_{i,j}^{l,s}-J^{l,s}_{\rm sr} \delta(z_j)&=&
J^{l,s}\sum_{i}\frac{1}{|{\bf r}_i-{\bf r}_j|^{d+\sigma_s}}  =
J^{l,s}\sum_{r_1=0}^{\infty}\sum_{r_2=-\infty}^{\infty}
\cdots\sum_{r_d=-\infty}^{\infty}
\frac{1}{\left[(z_j+r_1)^2+r_2^2+r_3^2+\cdots+
r_d^2\right]^{(d+\sigma_s)/2}}\nonumber \\
&\simeq& J^{l,s}\int_0^\infty dr_1\int_{-\infty}^\infty dr_2 \cdots
\int_{-\infty}^\infty dr_d \frac{1}{\left[(z_j+r_1)^2+
r_2^2+r_3^2+\cdots+r_d^2\right]^{(d+\sigma_s)/2}}\nonumber \\
&=&J^{l,s}\int_0^\infty dr_1\int_{0}^\infty dr
\frac{2\pi^{(d-1)/2}}{\Gamma\left(\frac{d-1}{2}\right)}
\frac{r^{d-2}dr}{\left[(z_j+r_1)^2+r^2\right]^{(d+\sigma_s)/2}}
\nonumber \\
&=&J^{l,s}\pi^{(d-1)/2}
\frac{\Gamma\left(\frac{1+\sigma_s}{2}\right)}
{\Gamma\left(\frac{d+\sigma_s}{2}\right)}
\int_0^\infty dr_1 (z_j+r_1)^{-\sigma_s-1}\nonumber \\
&=&J^{l,s}\pi^{(d-1)/2}
\frac{\Gamma\left(\frac{1+\sigma_s}{2}\right)}
{\sigma_s\Gamma\left(\frac{d+\sigma_s}{2}\right)} z_j^{-\sigma_s},
\label{Jls}
\end{eqnarray}
\end{widetext}
where $z_j \ge 1$ characterizes the distance of the particle $p_j$
from the boundary. We consider the fluid particles to be in the
region $0 \le z \le L$, where $L$ is the width of the film confined
between the two surfaces. Therefore, the partition function is
\begin{equation}
Z=\sum_{\{p^l\}}\exp\left[\beta \left( \sum_{j\in {\cal
L}}\mu_{j}p_{j}^{l}+{2} \sum_{i,j \in {\cal
L}}J_{i,j}^{l}p_{i}^{l}p_{j}^{l}\right)\right],
\end{equation}
i.e., the system is equivalent to one with a spatially varying
chemical potential $\mu_j=\mu-V_j$ acting on a particle $p_j$ at a
distance $z_j+1$, $0\le z_j \le L$, from the left boundary surface
and at a distance $(L+1-z_j)$ from the right one where $V_j\equiv
V(z_j)$ is the superposition
\begin{eqnarray}
\label{hj} V(z) &=& -\rho_s J^{l,s}_{\rm
sr}[\delta(z)+\delta(L-z)]\nonumber \\ && +v_s
\left[(z+1)^{-\sigma_s}+(L+1-z)^{-\sigma_s}\right],
\end{eqnarray}
where
\begin{equation}
\label{Avsd} v_s=-4\pi^{(d-1)/2}
\frac{\Gamma\left(\frac{1+\sigma_s}{2}\right)}
{\sigma_s\Gamma\left(\frac{d+\sigma_s}{2}\right)} \rho_s J^{l,s}.
\end{equation}
In the present study we choose $J^{l,s}_{\rm sr}$ such that
$\rho(0)=\rho(L)=1$, where $\rho(z)=\langle p_j^l \rangle$. This is
implemented by taking $J^{l,s}_{\rm sr}\rightarrow\infty$, which is
known as $(+,+)$ boundary conditions applied to the system under
consideration.

Note that in Eqs. (\ref{Jls}) and (\ref{hj}) only the leading
$z$-dependence of the substrate potentials has been retained. We
remark that there is some arbitrariness in what one calls the
distance $z$. One might measure it from the last substrate layer or,
say, from the midplane between the last substrate layer and the
first liquid layer. These different choices as well as taking into
account the actual discreteness of the substrate lattice in normal
direction (see the second line in Eq. (\ref{Jls})) lead to
additional terms decaying as $z^{-\sigma_s-1},\ z^{-\sigma_s-2}$,
etc. In the following we do not consider such subdominant
contributions.

One can easily determine the (leading) $L$-dependent contribution
$\Delta  \omega_{ex}^{(s,s)}$ in the excess grand canonical
potential $\omega_{ex}$ which is generated by the direct interaction
of the two substrate half-spaces a distance $L$ apart. According to
Eq. (\ref{hj}), one has
\begin{eqnarray}\label{fwall}
    \Delta  \omega_{ex}^{(s,s)} &=&
    \sum_{z=L+1}^\infty \rho_s v_s z^{-\sigma_s}
 \\ &\simeq & -4 \pi^{(d-1)/2}
\frac{\Gamma\left(\frac{1+\sigma_s}{2}\right)} {\sigma_s
\Gamma\left(\frac{d+\sigma_s}{2}\right)} \rho_s^2
J^{s}\int_{L+1}^{\infty}\!\!dz\,z^{-\sigma_s}\nonumber \\ &\simeq &
A_s(T) L^{-\sigma_s+1}, \nonumber
\end{eqnarray}
where $A_s$ is given in Eq.\,(\ref{as}). Also in this expression we
have retained only the leading $L$-dependent part.

\section{Finite-size behavior off criticality}
\label{offcritical}

\subsection{Finite-size contribution due to the long-ranged part
of  the fluid-fluid van der Waals interaction}

\label{Hamaker}

Away from $T_c$ the replacement of the fluctuating particle density
by its mean value, which is inherent to the present mean-field
approach, is a reliable approximation. This leads to the following
finite-size contribution ${\Delta \omega_{ex}^{(l)}}$ to the excess
grand canonical potential per area $A$ that is due to the
long-ranged tail of the fluid-fluid interaction
$\hat{w}=-4J^l/(1+|{\bf r}|^{d+\sigma})\; \theta(|{\bf r}|-1)$:
\begin{eqnarray}
\label{dat} \lefteqn{{\Delta \omega_{ex}^{(l)}}} \nonumber \\ &&
\hspace{-0.5cm} \equiv
  \lim_{A\rightarrow \infty} \frac{1}{2 A}
  \sum_{({\bf r}_\|,z), ({\bf r}'_\|,z')\in {\cal L}}
  \hat{w}({\bf r}_\|,z,{\bf
r}'_\|,z')\rho({\bf r}_\|,z)\rho({\bf r}'_\|,z') \nonumber \\
&\simeq& -{2}J^{l} \rho_b^2\sum_{{\bf r}_\parallel,z,z'}
  \frac{1}{1+\left[r_\parallel^2+(z-z')^2\right]^{(d+\sigma)/2}}\\
   &\simeq& -{4}J^{l} \rho_b^2 \frac{\pi^{(d-1)/2}}
   {\Gamma(\frac{d-1}{2})}
   \sum_{z \ne z'}\int_0^\infty\frac{r^{d-2}dr}
   {\left[r^2+(z-z')^2\right]^{(d+\sigma)/2}} \nonumber \\
   & \simeq & -{2}J^{l} \rho_b^2 \pi^{(d-1)/2}
   \frac{\Gamma(\frac{1+\sigma}{2})}{\Gamma(\frac{d+\sigma}{2})}
   \int_{1}^L dz \int_{1}^L
   dz'\  |z-z'|^{-(\sigma+1)}\nonumber \\
  &\simeq & A_l(T,\mu) L^{-\sigma+1}, \nonumber
\end{eqnarray}
where $A_l$ is defined in Eq.\,(\ref{at}) and the
above integration over $z$ has been performed under the restriction
$|z-z'|\ge 1$, which mimics the regularization of $\hat{w}$ at ${\bf
r}={\bf r}'$. In all above steps in Eq. (\ref{dat}) only the leading
$L$-dependent terms, which are not part of the bulk or the surface
contributions to the grand canonical potential, have been retained.

\subsection{Finite-size contributions of the long-ranged tails of
the substrate  potentials for $T \ll T_c$} \label{Pavel}

Away from $T_c$ the leading order finite-size contribution to the
grand canonical film potential due to the long-ranged tail of the
substrate potential $V(z)\equiv \kappa J^{l,s}_{\rm sr}\delta(z-1)+
v_s z^{-\sigma}$ of, say, the left wall is
\begin{eqnarray}\label{contrihs}
    \Delta \omega_{ex}^{(l,s)} &\simeq & \rho_b v_s \left(\sum_{z=1}^L
    z^{-\sigma_s}
    \right)\\ &\simeq &
    \rho_b v_s \int_1^L\!dz\,z^{-\sigma_s}=
    -\frac{\rho_b v_s}{(\sigma_s-1)}L^{-\sigma_s+1}, \nonumber
\end{eqnarray}
where we have again retained only the leading $L$-dependent part.

Taking into account the contribution of the second surface, which
renders a factor of 2 for the leading $L$-dependent part, we conclude that away from $T_c$ the
contribution of the tails of the substrate potentials to the
behavior of the Casimir force is
\begin{equation}\label{Casimirhs}
    \Delta f^{(l,s)}=-\beta\frac{\partial \Delta \omega_{ex}^{(l,s)}}{\partial
    L}=-2 \beta\rho_b v_s L^{-\sigma_s}=\beta (\sigma_s-1)A_{l,s},
\end{equation}
which is basically the result derived in Ref. \cite{ABUP91} (see Eq.
(6.11) therein) and is in full accordance with the findings of Ref.
\cite{MDB}. The constant $A_{l,s}$ is defined in Eq. (\ref{ah}).

\section{Order parameter profile in a film}
\label{dereqstate}

Since in the present study the external potential $V({\bf r})\equiv
V({\bf r}_\|,z)=V(z)$ depends  only on the coordinate normal to the
confining plates and there is no symmetry breaking in the lateral
directions, the density profile also depends only on $z$ so that
Eq. (\ref{eqstatestandard}) turns into the following equation for
the order-parameter profile:
\begin{eqnarray}\label{theeqwithG}
  \phi^*(z) &= & \tanh\Bigg[\beta \sum_{z'}J_l(z-z') \phi^*(z')\nonumber \\
  && +
  \frac{1}{2}\beta (\Delta \mu-\Delta V(z))\Bigg],
\end{eqnarray}
where
\begin{equation}\label{Gdef}
J_l(z)\equiv \sum_{{\bf r}_\|'}J({\bf r}_\|-{\bf
r}_\|',z)=\sum_{{\bf r}_\|}J({\bf r}_\|,z).
\end{equation}
Measuring distances in terms of the lattice spacing and taking the
interaction of the form (see Eq. (\ref{Jfluid}))
\begin{eqnarray}
J({\bf r})& \equiv & J^l_{\rm sr}\,[\delta(|{\bf r}|)+\delta(|{\bf
r}|-1)]+\frac{J^l}{1+|{\bf r}|^{d+\sigma}}\; \theta(|{\bf r}|-1)
\nonumber \\
&=& (J^l_{\rm sr}-J^l)\, \delta(|{\bf r}|)+(J^l_{\rm sr}-\frac{1}{2}
J^l)\, \delta(|{\bf r}|-1) \nonumber \\ && +\frac{J^l}{1+|{\bf
r}|^{d+\sigma}} \label{inter}
\end{eqnarray}
 one can
further simplify the sum on the right-hand side of Eq.
(\ref{theeqwithG}). Using the identity \cite{B89} (see also Ref.
\cite{danchev})
\begin{equation}\label{id}
\frac{1}{1+y^\alpha}=\int_0^\infty dt \ \exp(-y t)\; t^{\alpha-1}\
E_{\alpha,\alpha} (-t^\alpha),
\end{equation}
where
\begin{equation}
\label{ML} E_{\alpha,\beta} (y)=\sum_{k=0}^\infty
\frac{y^k}{\Gamma(\alpha k +\beta)}, \qquad \alpha>0,
\end{equation}
 are the Mittag-Leffler functions, the sum over ${\bf r}_\|$ in Eq.
 (\ref{Gdef}) can be rewritten as
 \begin{eqnarray}
 \label{Jlz}
 J_l(z) &=&
[(J^{l}_{\rm sr}-J^l)+ 2(d-1)(J^{l}_{\rm sr}-\frac{1}{2}J^l)]\;
\delta(z)
\\ && +(J^{l}_{\rm sr}-\frac{1}{2}J^l)[\delta(z-1)+\delta(z+1)]+
 J_{d,\sigma}^l(z) \nonumber
 \end{eqnarray}
 with
 \begin{eqnarray}\label{Gds}
J_{d,\sigma}^l(z) &=& J^l \int_0^\infty dt \ t^{(d+\sigma)/2-1}\
E_{\frac{d+\sigma}{2},\frac{d+\sigma}{2}}(-t^{(d+\sigma)/2}) \times
\nonumber \\ &&
\left(\sum_{{\bf r}_\|}e^{-t{\bf r}_\|^2}\right)
e^{-tz^2}.
 \end{eqnarray}
The main advantage of the above form is that it factorizes the
summation over the components of ${\bf r}_\|$. Using the Poisson
identity
\begin{equation}\label{Pi}
    \sum_{r=-\infty}^\infty e^{-t
    r^2}=\sqrt{\frac{\pi}{t}}\sum_{n=-\infty}^\infty e^{-\pi^2
    n^2/t}
\end{equation}
one has
\begin{eqnarray}
\label{sumgenfinal} \lefteqn{\sum_{z'=0}^{L} J_{d,\sigma}^l(z-z')
\phi(z')= J^l \Bigg\{c_d \ \phi(z)} \\ && + \sum_{z'=0 \atop z\ne
z'}^{L} \Big[g_{d,\sigma}(|z-z'|)+g_{d,\sigma}^{nn}(|z-z'|)
\Big]\phi(z')\Bigg\}, \nonumber
\end{eqnarray} where
\begin{equation}
c_d = \sum_{{\bf r}_\|}\frac{1}{1+{\bf r}_\|^{d+\sigma}}
\end{equation}
is a constant depending on $d$ and $\sigma$. The function
\begin{eqnarray}
g_{d,\sigma}(a) &=& \int_0^\infty dt \
\left(\frac{\pi}{t}\right)^{(d-1)/2} t^{\frac{d+\sigma}{2}-1}\times
\nonumber \\ &&
E_{\frac{d+\sigma}{2},\frac{d+\sigma}{2}}\left(-t^{\frac{d+\sigma}{2}}\right)
e^{-t a^2} \label{gds_def}
\end{eqnarray}
reflects the contribution of the long-ranged van der Waals tails of
the interaction, and the function
\begin{eqnarray}
\lefteqn{g_{d,\sigma}^{nn}(a)= \int_0^\infty dt \
\left(\frac{\pi}{t}\right)^{(d-1)/2} t^{\frac{d+\sigma}{2}-1}
\times} \nonumber \\ &&
E_{\frac{d+\sigma}{2},\frac{d+\sigma}{2}}\left(-t^{\frac{d+\sigma}{2}}\right)\sum_{{\bf
n} \in {\mathbb Z}^{d-1} \atop {\bf n}\ne {\bf 0}}e^{-\pi^2{\bf
n}^2/t-t a^2}
\end{eqnarray}
concerns mainly the coupling between neighboring sites in adjacent
layers (see below). Here ${\bf n}\in {\mathbb Z}^{d-1}$ is a
$(d-1)$-dimensional vector with integer components because all
lengths are measured in units of the lattice spacing. It is easy to
show that $\max_{t}\exp(-\pi^2 {\bf n}^2/t+t(z-z')^2)$ is attained
at $t=\pi |{\bf n}|/|z-z'|$ and equals $\exp({-2\pi |{\bf n}|
|z-z'|})$. Due to this exponential decay in the last term of Eq.
(\ref{sumgenfinal}) one is able to take into account only the terms
with $|{\bf n}|=1$ and $|z-z'|=1$, which amounts to the
approximation $g_{d,\sigma}^{n,n}(a)\simeq g_{d,\sigma}^{nn}(\pm
1)\equiv \hat{c}_{d,\sigma}^{nn}$. It is straightforward to take
into account additional, smaller terms  corresponding to $|{\bf
n}|=2, 3, \cdots$ and $|z-z'|=2, 3, \cdots$, but it turns out that
already the contributions stemming from $|{\bf n}|=1$ and $|z-z'|=1$
are numerically very small. Thus the size dependent contributions
due to the last term on the right-hand side of Eq.
(\ref{sumgenfinal}) are exponentially small in $L$.

\begin{widetext}
For $d=\sigma=4$ the corresponding Mittag-Leffler function can be
expressed as
\begin{eqnarray}\label{ML4}
E_{4,4}(-t^4) =
\frac{1}{\sqrt{2}t^3}\Bigg[\cosh\left(\frac{t}{\sqrt{2}}\right)
\sin\left(\frac{t}{\sqrt{2}}\right) -
\cos\left(\frac{t}{\sqrt{2}}\right)\sinh\left(\frac{t}{\sqrt{2}}\right)
\Bigg].
\end{eqnarray}
For $a>1/\sqrt[4]{2}$ this leads to the following representation of
$g_4(a) \equiv g_{4,4}(a)$ (see Eq. (\ref{gds_def})):

\begin{eqnarray}
g_4(a)&=& \pi^{3/2}\int_0^\infty dt\
t^{3/2}E_{4,4}(-t^4)\exp\left[-t
a^2 \right] \nonumber \\
&= & \frac{\pi^2}{2^{3/4}}\left \{   \left[
1+\left(\sqrt{2}a^2+1\right)^2
\right]^{1/4}\left[\sin\left(\frac{1}{2}{\rm arccot}[\sqrt{2}a^2+1]
\right) -\cos\left(\frac{1}{2}{\rm
arccot}[\sqrt{2}a^2+1] \right)\right] \right. \nonumber \\
& & +\left. \left[ 1+\left(\sqrt{2}a^2-1\right)^2
\right]^{1/4}\left[\sin\left(\frac{1}{2}{\rm arccot}[\sqrt{2}a^2-1]
\right) +\cos\left(\frac{1}{2}{\rm arccot}[\sqrt{2}a^2-1]
\right)\right]\right \} \label{g4ap}
\end{eqnarray}
and the  equation for the order parameter profile becomes
\begin{equation}
\label{d4em} {\rm arctanh}\left[ \phi^*(z)\right]= \frac{1}{2}\beta
(\Delta \mu-\Delta V(z))+ K \left\{a_4 \phi^*(z) + a_4^{nn}
\left[\phi^*(z+1)+\phi^*(z-1)\right]+\lambda \sum_{z'=0 \atop |z'-
z|\ge 2}^{L}g_4(|z-z'|)\phi^*(z')\right\},
\end{equation}
\end{widetext}
where $K=\beta J^{l}_{\rm sr}$, and $\lambda=J^{l}/J^{l}_{\rm sr}$.
In Eq. (\ref{d4em}) one has
\begin{equation}\label{a3}
a_4=7+\lambda (c_4-4), \qquad a_4^{nn}=1+\lambda
(c_4^{nn}-\frac{1}{2}),
\end{equation}
where
\begin{equation}\label{c2}
c_4=\sum_{{\bf n}\in {\mathbb Z}^3}\frac{1}{1+|{\bf n}|^8}\simeq
4.900
\end{equation}
and
\begin{eqnarray}\label{c2nn}
c_4^{nn}&=& g_4(1)+\hat{c}_{4,4}^{nn}\simeq  1.015+0.013 \approx
1.028.
\end{eqnarray}
This shows that the contribution of $\hat{c}_4^{nn}$ to $c_4^{nn}$
is ca. $1\%$.
\begin{figure}[h]
\includegraphics[angle=0,width=\columnwidth]{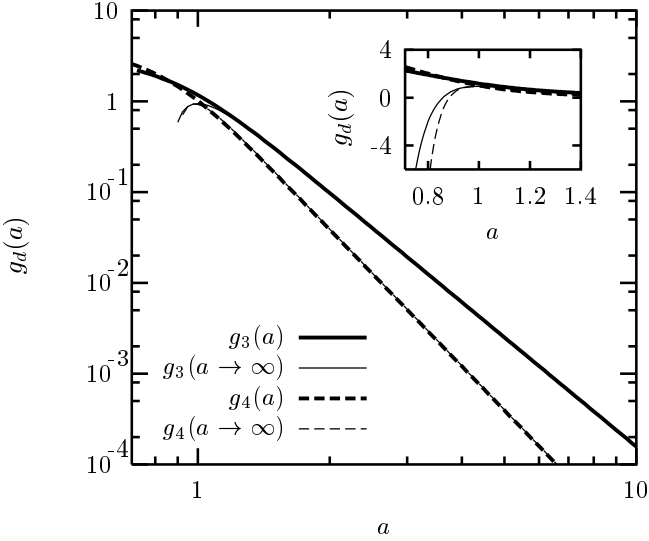}
\caption{The behavior of the function  $g_d(a)$ and its asymptote
for $d=\sigma=3$ and $d=\sigma=4$. \label{gplot}}
\end{figure}

For the asymptotic behavior one finds
\begin{equation}
\label{g4a} g_4(a\rightarrow\infty)=\frac{\pi^2}{8}a^{-5}-\frac{33
\pi^2}{1024}a^{-13}+ O(a^{-21}).
\end{equation}
By setting $g_4 \equiv 0$,  Eq.\,(\ref{d4em}) reduces to the
well-known form of the mean-field theory for systems with
short-ranged fluid-fluid forces.  (Actually in our case setting
$g_4\equiv 0$ corresponds to  a system with short-ranged
interactions in $z$-direction and long-ranged ones within the planes
perpendicular to $z$; only setting $\lambda=0$ or, equivalently,
$J^l=0$ reduces the model to one with strictly short-ranged
fluid-fluid interactions.) The standard Ginzburg-Landau equation
follows from taking the expansion ${\rm arctanh}(\phi\rightarrow 0)=
\phi + \phi^3/3 +O(\phi^5)$. A continuum version of the equation
follows from the replacement $\phi(z-1)+\phi(z+1)\rightarrow
2\phi(z)+\phi''(z)$. Obviously such a continuum version can be
constructed also for a system with long-ranged forces by taking into
account the function $g_4(a)$ (which then has to be considered as a
function of a continuous variable).
This exploits the fact that the function $g_4(a)$ is well defined
for $a \ge 0$ (provided that for $0<a<1/\sqrt[4]{2}$ one takes the
principal values of ${\rm arccot}$ to be in the interval $(0,\pi)$)
and not only for $a\ge 1$, as actually needed for the lattice
formulation of the theory. Therefore in the continuum formulation of
the theory the integration can be extended to the full interval
$z\in [0,L]$. This does not change the algebraic asymptotes of the
density profiles. Thus in the continuum case the equation for the
density profile in the presence of long-ranged interactions is an
integro-differential equation:
\begin{eqnarray}\label{eqcont}
\lefteqn{\phi^*(z)+\frac{1}{3}(\phi^{*}(z))^3=\frac{1}{2}\beta
(\Delta \mu-\Delta V(z))} \nonumber \\ &&+K\Bigg\{a_4 \phi^*(z) +
a_4^{nn} \left[2 \phi^*(z)+\frac{d^2
\phi^*(z)}{dz^2}\right]\nonumber \\ && +\lambda
\int_{0}^{L}g_4(|z-z'|)\phi^*(z')dz' \Bigg\}.
\end{eqnarray}
The behavior of the function $g_4(a)$ as well as its asymptote (Eq.
(\ref{g4a})) are shown in Fig.\,\ref{gplot}. Equation (\ref{eqcont})
has to be augmented by the boundary conditions $\phi^*(z\rightarrow
0)=\phi^*(z\rightarrow L)=\infty$.

Note that, according to Eqs. (\ref{d4em}) (or (\ref{eqcont}) in the
continuum case) and (\ref{hj}), the only {\it explicit} $L$
dependence in the behavior of the order parameter profile enters via
the asymptotes of the functions $V(z)$ and $g_4(|z|)$ for $|z| \gg
1$. They will then lead to contributions  of the order of
$L^{-\sigma}$
 in the behavior of $\rho$ away from $T_c$, i.e., there $\rho \simeq
\rho_b +O(L^{-\sigma})$. The last result implies that the entropy
term in Eq. (\ref{freeenergyfunctionalstarting}) in this regime will
produce contributions to $\omega_{\rm ex}$ of the order of
$L^{-\sigma}$. Such contributions can then be neglected in
comparison with the ones described by $A_l$ (see Eq. (\ref{dat})),
$A_s$ (see Eq. (\ref{fwall})), and $A_{l,s}$ (see Eq.
(\ref{Casimirhs})). Obviously this argument is not specific for
$d=4$ but is  generally valid. Finally, we repeat that in the case
$d=\sigma=3$ also an analytical expression for $g_3(z)$ can be
derived \cite{DR2006}:
\begin{eqnarray}
 g_3(a)&=&\frac{\pi}{3}\left[\sqrt{3}\ \arctan{
\frac{\sqrt{3}}{2a^2-1}}-\ln\left(1+\frac{1}{a^2}\right) \right.
\nonumber \\ && \left.+
\frac{1}{2}\ln\left(1-\frac{1}{a^2}+\frac{1}{a^4}\right)\right].
 \label{g3}
\end{eqnarray}
The behavior of this function as well as its asymptote  are also
shown in Fig.\,\ref{gplot}.

\flushright


\begin{thebibliography}{99}
\bibitem{rem4} For the discrete delta function $\delta(x)$ we use
the definition $\delta(x)=1$ if $x=0$, and $\delta(x)=0$ if $x\ne
0$.
\bibitem{rem1} If one uses the notations of Ref.
\cite{ABUP91}, i.e.,  $4 J^l \rightarrow A, d+\sigma\rightarrow
n$, for $d=3$ this term can be rewritten as $- A
\frac{2\pi}{(n-2)(n-3)(n-4)}L^{-\sigma}$, which is identical to Eq.\,(6.11)
in Ref. \cite{ABUP91}.
\end{thebibliography}
\end{document}